\documentclass[amsmath,amssymb,pre,notitlepage]{revtex4-1}
\pdfoutput=1
\usepackage{bm}
\usepackage{booktabs}

\usepackage{cancel}
\usepackage[normalem]{ ulem }
\usepackage{soul}
\usepackage[dvipsnames]{xcolor}
\usepackage{soul}

\usepackage{physics}
\usepackage{graphicx}
\graphicspath{ {images/} }
\usepackage{xfrac} 
\usepackage{float} 
\usepackage{stmaryrd} 
\usepackage{enumitem}
\usepackage{svg}
\usepackage{ragged2e}
\usepackage{caption}
\usepackage{wrapfig}
\sethlcolor{yellow}

\usepackage[hidelinks]{hyperref}

\begin{document}
 \title{On the side-effects of compromising: coupling agents' heterogeneity with network effects on a bounded confidence opinion dynamics model.}

\author{R\'{e}mi Perrier}
 \email{remi.perrier@cyu.fr}
 \affiliation{Laboratoire de Physique Th\'{e}orique et Mod\'{e}lisation, UMR-8089 CNRS, CY Cergy Paris Universit\'{e}, France}
 
 \author{Hendrik Schawe}
 \email{hendrik.schawe@cyu.fr}
 \affiliation{Laboratoire de Physique Th\'{e}orique et Mod\'{e}lisation, UMR-8089 CNRS, CY Cergy Paris Universit\'{e}, France}

 \author{Laura Hern\'{a}ndez}
 \email{laura.hernandez@cyu.fr}
 \affiliation{Laboratoire de Physique Th\'{e}orique et Mod\'{e}lisation, UMR-8089 CNRS, CY Cergy Paris Universit\'{e}, France}

 \email{all@}

\begin{abstract}
We present an extensive study of the joint effects of  heterogeneous social agents and their heterogeneous social links in a bounded confidence opinion dynamics model.
 The full  phase diagram of the model is explored for  two different network's topologies and compared to two opposed extreme  cases:  on one hand heterogeneous agents which constitute a  mixed population and on the other, their interactions are modeled by a  lattice. The results show that when  agents prone to compromising coexist with close minded ones, the steady state of the dynamics shows coexistent phases. In particular, unlike the homogeneous case in networks, or the heterogeneous case in fully mixed population, it is possible that the society ends up in consensus around one extreme opinion. Moreover,   during the dynamics, the consensus may be overturned from one extreme to the other of the opinion space. We also show that the standard order parameter, the normalized average size of the largest opinion cluster, may be misleading in this case,  as it hides the existence of these phases. The phase where the  opinion of the society is overturned does not require the presence of  agents with special characteristics, (stubborn, extremists, etc.); it results from the interplay of agents which have agreed on an extreme opinion with the remaining group that holds the opposite one. Among the former, some may be  prone to compromising  with other agents which are out of the majority group, these agents, according to their location in the network,  may act like bridges between the two groups and slowly attract the whole society to the other extreme.
\end{abstract}

\nopagebreak
\maketitle

\section{Introduction}\label{intro}

Nowadays, the challenge of theoretical studies on opinion dynamics is to consider realistic properties that  were absent from the first stylized models, in order to explore their role on the outcomes of the dynamics~\cite{Castellano2009Statistical, Sirbu2017Opinion}.
 These original stylized models have nevertheless been successful in revealing the leading role played by the microscopic interactions at the level of the agents,  often inspired on Social Influence  Theory~\cite{deutsch1955study,latane1981psychology},  on the transition from a fragmented opinion to a  consensus or polarized phase.  Amongst the most famous  results, one can cite    the low threshold for consensus in bounded confidence models~\cite{deffuant2000mixing,hegselmann2002opinion}, the intrinsic dynamical character of segregation~\cite{schelling1978sorting}, or the unexpected coexistence of multicultural societies, in spite of the presence of local homogenizing interactions~\cite{Axelrod1997}. 

The ultimate goal in this  domain would be to fill the gap
 between the numerous theoretical models and the growing number of empirical studies derived from the  increasing amount of data that is nowadays easily accessible, mainly due to the widespread usage of social networks. However,  a model that integrates as many details of a real society as possible will not be explanatory, due to the large number of uncontrolled parameters. Instead, the safe road to the comprehension of the behaviour of   realistic models systematically study   the role that each of the main properties characterizing real social systems, play  on the outcomes of the dynamics. 
 
 An unavoidable aspect to be considered in the study of real  social systems is heterogeneity. Social actors and their interactions are intrinsically heterogeneous, displaying idiosyncratic  characteristics that may evolve -when they do- at a very slow timescale with respect to that of the opinion formation. 
 On this line,  several works have extended existing models to include some realistic properties of social interactions. 
An important  extension takes into account that not all the agents interact with the same number of people and certainly  not with the whole population, as in the fully mixed population hypothesis, or \textit{mean-field approach}; instead,  social interactions are modeled by a network of social agents connected  by links which  may be endowed with the properties of social ties observed in real life~\cite{fortunato2005consensus, fortunato2004universality,schawe2021bridges}. 

Another source of heterogeneity is rooted in the fact each individual has personal characteristics that lead to different reactions to the same stimuli. This is usually modeled by a quenched disorder distributed over the population. Like in  physical systems, quenched disorder generally  leads  to  important modifications of the  results established for models without disorder,  even in the simple  mixed population scenario~\cite{lorenz2003opinion,lorenz2005stabilization,Lorenz2010Heterogeneous,Kou2012multi,schawe2020open,schawe2020cost}. 

Moreover in real life, individuals interact with peers who  belong to different spheres of the society, like work, friends, family, etc. in such way that a modification of their opinion due to the influence of neighbours in one social environment is automatically transferred to  the others. This aspect can be  naturally treated by  multiplex network models of the society, where each layer contains the same agents, but where the interactions are different from one layer to another~\cite{Amato_2017,PhysRevE.104.064304,PhysRevE.99.062311}. 

Very recent works address the fact that not all the social interactions can be decomposed into pairwise ones, because in some cases, the interaction takes place at the level of a group. Hyper-graphs, where the individuals are linked together by an hyper-edge, constitute the ideal tool to model these situations~\cite{lambiotte2019networks,hickok2021bounded,schawe2022higher}.

The full phase diagram of the heterogeneous version of the Hegselmann-Krause (HK) opinion dynamics model~\cite{hegselmann2002opinion} in a mixed population setting has recently been extensively studied~\cite{schawe2020open}. 
 The HK model is a \textit{bounded confidence} model based on two main ingredients of social interaction, \textit{homophily}, which designates the fact that "likes attract" and \textit{social influence}, which correspond to the fact that humans have the tendency to be  influenced by those who are similar to them, while they disregard the opinions of others whose opinion is far from their own. 
 
 In the HK  model, opinions are allowed to vary continuously, and therefore a continuous dynamical variable is associated to each agent representing the evolution of its opinion in time. Social agents  interact only with those peers whose opinion differ from their own in less than a parameter value, $\varepsilon$, known as \textit{confidence}, materializing the open-mindedness of the society. The dynamics is a synchronous operation, where all the agents update their opinion to the average opinion of  the set of agents they interact with, and  their own. 
The corresponding heterogeneous version considers that, instead of having a single confidence value for the whole population,  the confidence of social actors is an idiosyncratic characteristic:  while some are very  \textit{open- minded}, ready to interact even  with unlike peers, others are   \textit{close-minded} and  interact only with those holding an opinion very similar to theirs.  As there is no way to parameterize the open-mindedness with real data,  a whole exploration of the phase diagram is necessary. The main result is that, paradoxically, introducing additional more open-minded agents in a society that contains some close-minded ones, does not necessarily lead to   a consensus phase.  

On the other hand, it has been shown by extensive finite size analysis, that when the agents are homogeneous (all holding the same value of the confidence), the introduction of  heterogeneity in the social ties, constraining them by a random network, makes the critical value of the confidence that allows to establish consensus, $\varepsilon_c$, to decrease with the system size, vanishing in the thermodynamic limit. This effect is absent in fully mixed populations and lattices~\cite{schawe2021bridges}.

Real societies mix these two sources of heterogeneity, as they are composed of  heterogeneous agents who are bound by a network of social contacts with whom they mainly interact. 
In this work we present a new step towards the understanding of real systems, by  integrating both aspects in a single model: heterogeneous agents with heterogeneous connections. To do so, we consider a HK model where each agent,$i$, has its own value of confidence $\varepsilon_i$ and where their potential interactions are constrained by a network.  Networks of  different topology are studied for comparison, including lattices and the complete network setting.  Our results show that the combination of  both types  of  heterogeneity gives rise to non trivial  phenomena; in particular, the presence of coexisting phases where the society shows either a weak consensus around the middle opinion (i.e; $x \approx 0.5$), or where the society shows a strong consensus around an extremist opinion  (i.e; $x \approx 1 $ or  $x \approx 0 $ ). This work also reveals that a few open-minded agents may be able to overturn the  majoritarian opinion from one extreme to the other.  By comparing with previous  results, we identify the  role of each aspect of heterogeneity in the observed outcomes of the dynamics, and we reveal new  phenomena that are the result of their joint action.

\section{Results}\label{results}

We have performed extensive simulations of the heterogeneous HK model where   the connections among the agents are constrained by different types of network and lattices.

Each  society is characterized by a set of \textit{quenched} variables representing the confidence of each agent $i$,  $\varepsilon_i$,   uniformly distributed  in the interval $[\varepsilon_l,\varepsilon_u]$, representing the range of 
open-mindedness of the society. Moreover, social ties are   modeled by a fixed network (lattice), where the vertices represent the agents and the links restrict their possible interactions to their neighbours in the network (lattice).  

The opinion of each agent $i$ at discrete time $t$ is represented by the  \textit{dynamic} variable $x_i(t) \in [0,1]$. Each agent $i$  interacts  with its neighbours in the network (lattice), provided that their difference of opinion is smaller than $\varepsilon_i$. It should be noted that the heterogeneity in the confidence values introduces an asymmetry in the interaction with respect to the homogeneous version of the  model~\cite{hegselmann2002opinion,fortunato2004universality,fortunato2005consensus,schawe2021bridges} (see Methods).

For each simulation, the agents'  initial opinions $x_i(t=0)$ are uniformly distributed in the interval $[0,1]$. Unless stated otherwise, averages are computed over $1000$ samples of each society, except for the phase diagrams, where due to the detailed covering of the phase space, averaging was limited to $100$ samples.

\subsection{Phase diagram of the model considering different network topologies}\label{topo}

\begin{figure}[h]
	\centering
	\includegraphics[width=0.8\linewidth]{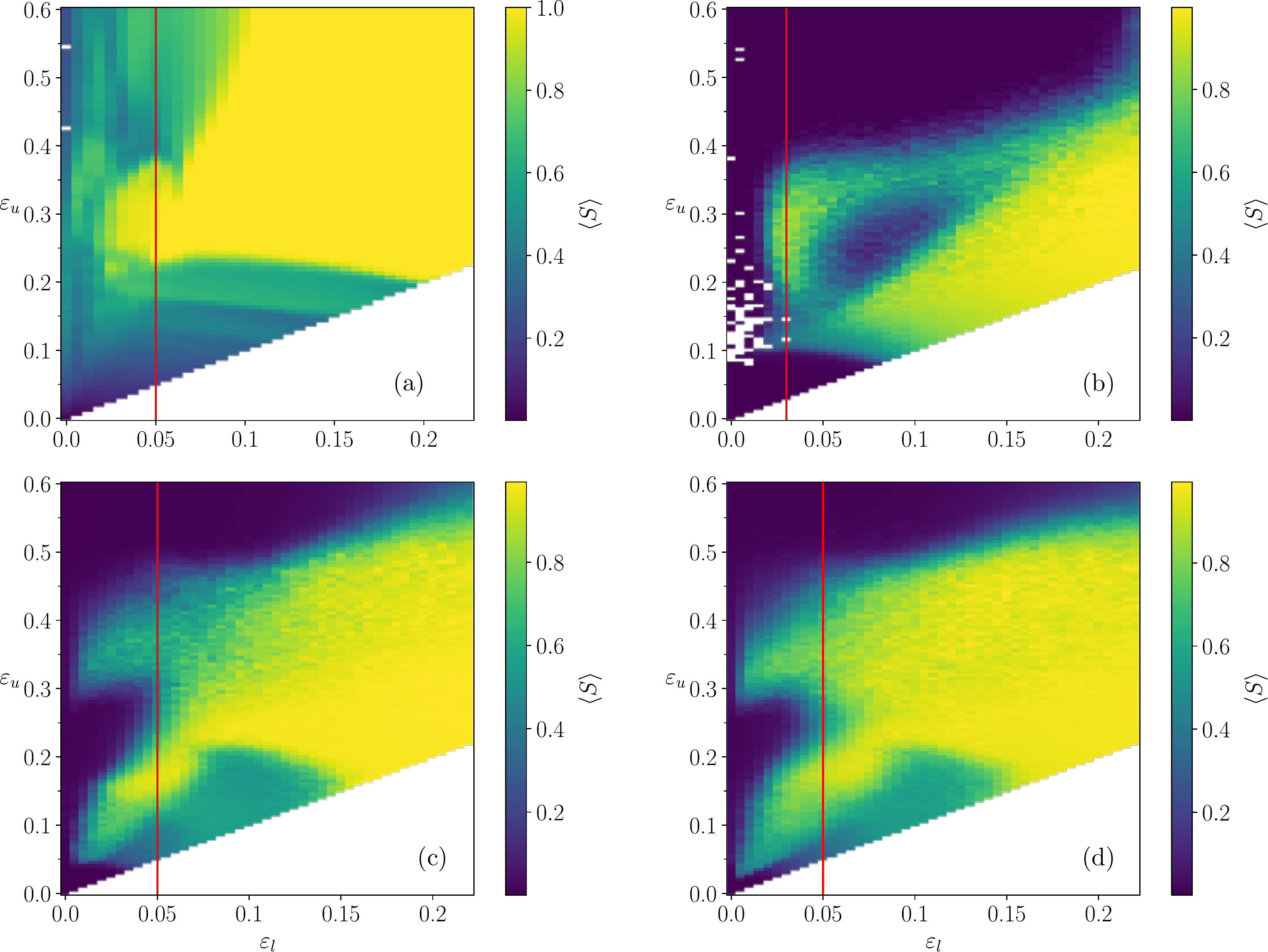}
	\captionsetup{justification=Justified}    
	\caption{Phase diagram for the heterogeneous HK model in systems of  $N=16384$ agents. The color map represents the average normalized size of the largest cluster $\langle S \rangle$ at the steady state of the system. (a): Fully connected network (taken with the permission of the authors) from~\cite{schawe2020open}. This depicts the average over 1000 realizations for each of the 2923 points, with a resolution of $6.25\times 10^{-3}$ on $\varepsilon_l$ and $\varepsilon_u$.  (b): Square lattice including up to third neighbour interactions, so that the coordination is $k=12$. (c): Erdős–Rényi network with $\langle k \rangle=10$ and  (d): Barabási-Albert network with $\langle k \rangle=10$. Panels (b-d) show an extensive exploration of the phase space with  4455 different points, with a resolution of $5\times 10^{-3}$ on $\varepsilon_l$ and $\varepsilon_u$. In this case,  averages have been taken over 100 realizations. White pixels correspond to societies where we did not compute the average to avoid selection bias because of a few realizations which did not converge in reasonable computing time (see section~\ref{cycle} below). Red lines denote the $\varepsilon_l = 0.05$ and $\varepsilon_l = 0.03$ slices to be discussed in detail later.}
	\label{S_comparo_topo}
\end{figure}

 The color map of Fig.~\ref{S_comparo_topo} represents the phase diagram of the system based on standard the order parameter, usually  considered in opinion dynamics studies: the average normalized size of the largest opinion cluster at the steady state, $\langle S \rangle$, of the HK dynamics for different societies. Each panel corresponds to a different topology of the interactions' network and, inside each panel, each pixel corresponds to  $\langle S \rangle$, calculated by averaging over 100 realizations of a given  society. A given society is defined by the coordinates $(\varepsilon_l, \varepsilon_u)$ giving the interval   from which the confidence parameters  of the   agents have been drawn, and by type of  network representing their potential interactions.  Clearly, the resulting  phase diagram strongly depends on the interplay of the heterogeneity of the agents and the topology of the network  that constrains the possible interactions. 

Figure~\ref{S_comparo_topo}(a),  corresponds to the case of the mixed population studied in Ref.~\cite{schawe2020open} and is given as a reference. The effects exclusively related to the network constraining  the interactions  appear along the diagonals which represent homogeneous systems:  the critical value to reach consensus, $\varepsilon_c$, diminishes in all networked societies with respect to the fully mixed population~\cite{schawe2021bridges}.  On the other hand, the effect of heterogeneity is also observed in all the panels: in societies with low confidence agents (low $\varepsilon_l$), the fact of adding more confident agents does not always enhance consensus. The strongest effect of combining  both properties (heterogeneous agents and networked societies) can be observed along the red vertical lines in Fig.~\ref{S_comparo_topo}: the order parameter $\langle S \rangle$ falls down to zero in networked societies where  polarization was observed in the mixed population. (See Supplementary material for the phase plots of lattices of different coordination numbers. 
A 360° visualization of the 3D scatter plots of the largest cluster size for each realization in the steady state, leading to the average represented in Fig.~\ref{S_comparo_topo}, can be found in the Supplementary Material I.A.)

In order to understand the differences with the  mixed population case, we study in detail systems  located in the region where a non monotonous behaviour was found in Ref.~\cite{schawe2020open}, marked with a vertical  red line in  Fig.~\ref{S_comparo_topo}.

\begin{figure}[h]
	\centering
	\includegraphics[width=\linewidth]{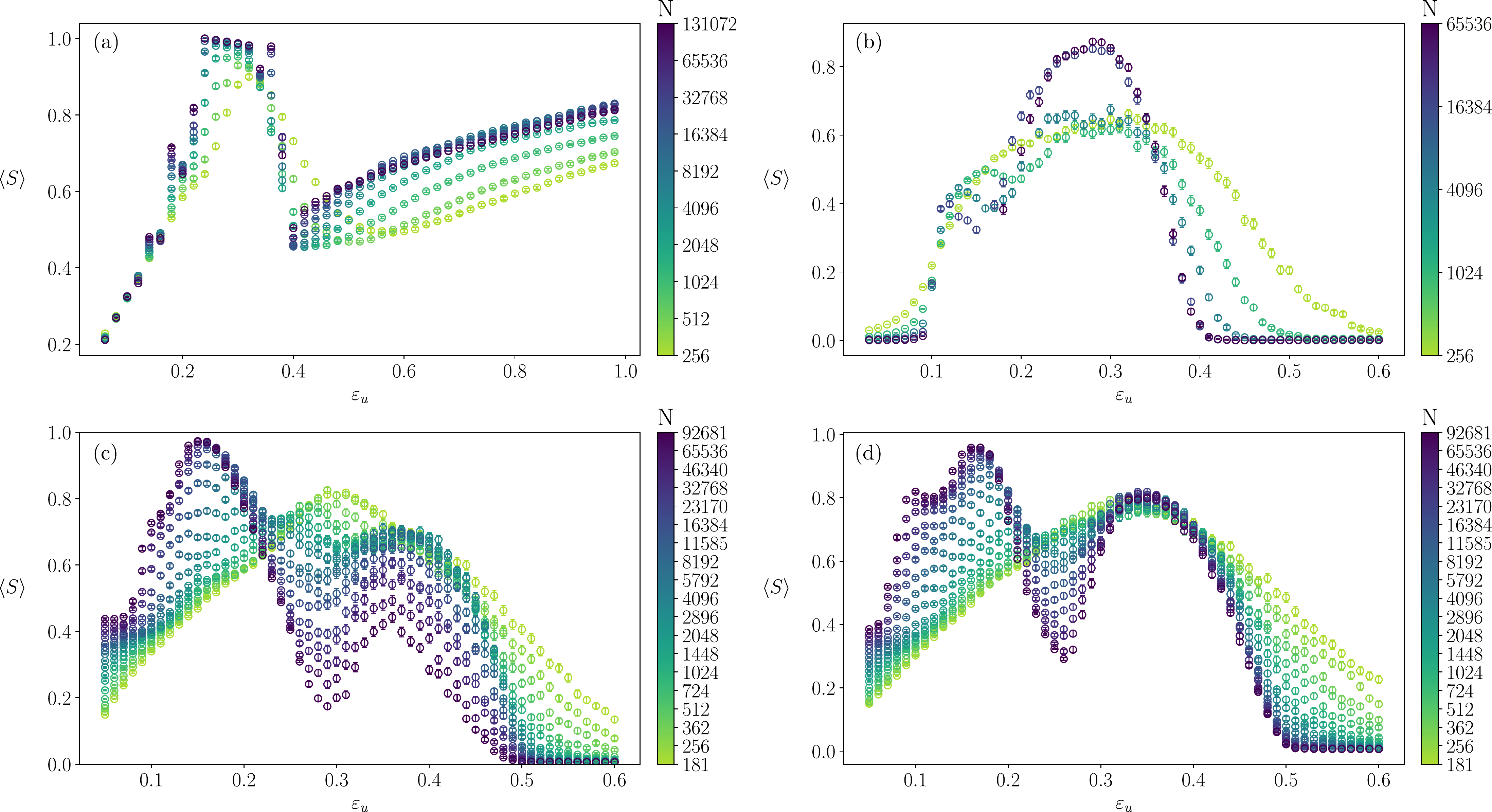}
	\captionsetup{justification=Justified}  
	\caption{Average relative size of the largest cluster, $\langle S \rangle$,   over 1000 realizations as a function of $\varepsilon_u$ for fixed $\varepsilon_l$ corresponding to the red vertical line of Fig.~\ref{S_comparo_topo}. The color map indicates different considered system sizes. (a) Fully connected network (data taken with the permission of the authors from~\cite{schawe2020open}), $\varepsilon_l=0.05$. (b) Square lattice including up to third neighbour interactions, $\varepsilon_l=0.03$. (c) Erdős–Rényi network of $\langle k \rangle=10$, $\varepsilon_l=0.05$. and (d) Barabási-Albert network of $\langle k \rangle=10$, $\varepsilon_l=0.05$.}
	\label{sizes_effects}
\end{figure}

Figure~\ref{sizes_effects} shows the order parameter,  $\langle S \rangle$, as a function of $\varepsilon_u$, for fixed $\varepsilon_l$,  different system sizes, and different networks' topology (the behaviour of other quantities, is shown in the  Supplementary Material). Strong size effects are observed and, although the global shape of the curves seems to stabilize as $N$ increases, the non monotonous behaviour of the order parameter  prevents us from finding a single scaling form. A similar  qualitative behaviour is observed for both random networks except for the behaviour of the  second peak  with $N$.

Figure~\ref{fit_second_peak} compares the scaling behaviour of the location and the height of this second peak, showing that it vanishes for large $N$ for the ER network, while it stays for the BA topology. In both cases, the position of the second peak does not appear to shift with the system size and stabilises around $\varepsilon_u = 0.36$ and  $\varepsilon_u = 0.35$  respectively.

\subsection{The  coexisting phases}

The previous results show that, unlike  the mixed population case, when the potential  interactions are constrained by a network,  $\langle S \rangle \longrightarrow 0$ for large values of $ \varepsilon_u$, which  seems counter-intuitive. In Ref.~\cite{schawe2020open} it was shown that, for low values of $\varepsilon_l$, as the fraction of  open minded agents  increases,  consensus turns into polarisation. This is caused by the  fast convergence of the very open minded agents to a  central opinion, who leave  aside the close minded ones. The latter, becoming unable to interact with the majoritarian strand, form a secondary cluster of relative importance, leading to a polarized opinion state.  Here, this explanation does not hold as there is no dominant cluster. In other words: the open minded ones do not form a large cluster either.
\pagebreak

\begin{wrapfigure}{r}{0.4\textwidth}
     \centering
     \includegraphics[width=\linewidth]{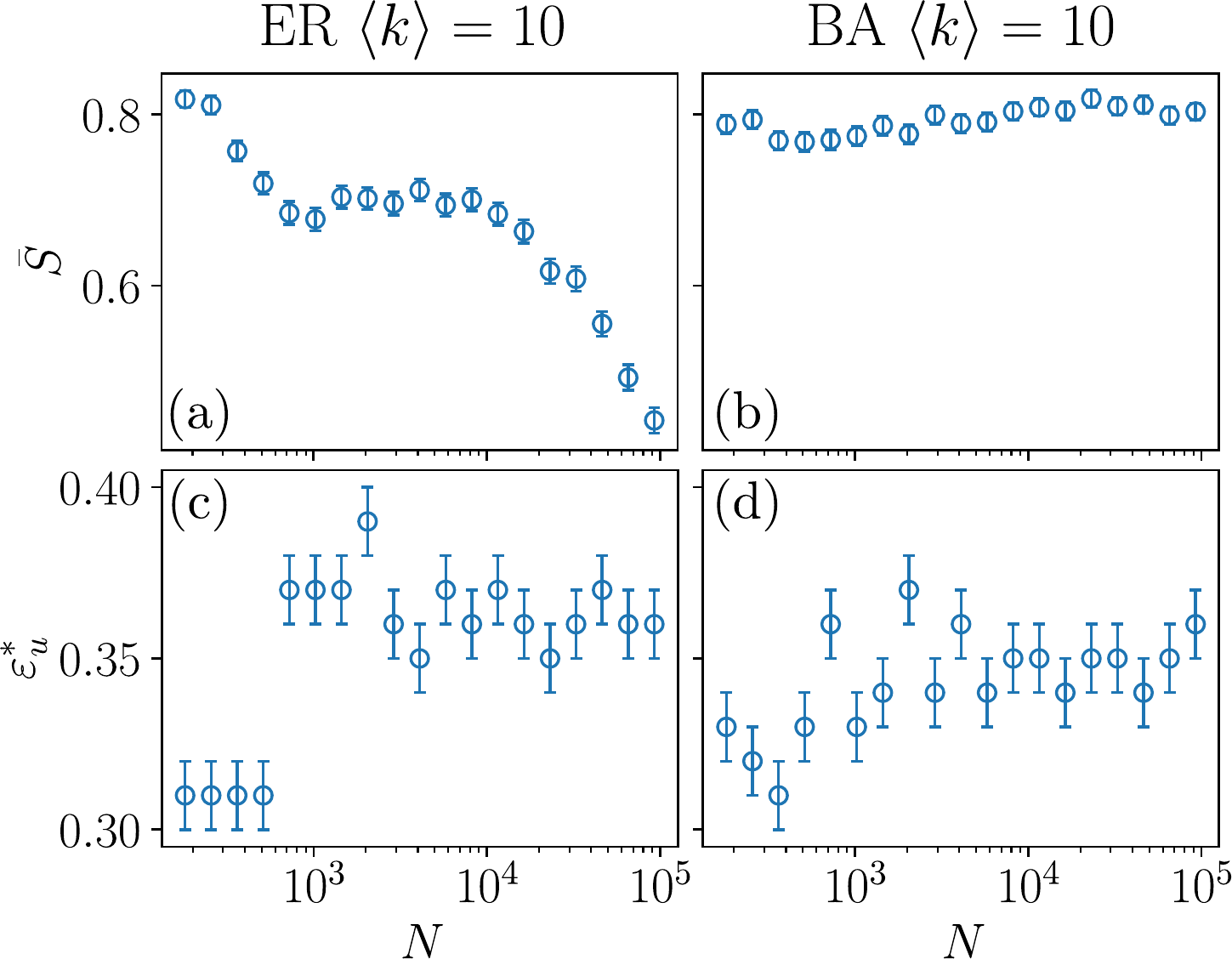}
     \captionsetup{justification=Justified}  
     \caption{Comparison of size effects on the second peak of the order parameter, $\langle S \rangle$, as a function of the system size for ER and BA interaction networks with $\varepsilon_l = 0.05$. (a) and (b) $\bar{S}$,  the height of the second peak, defined as the maximum value reached by $\langle S \rangle$ in the interval $\varepsilon_u\in[0.3, 0.5]$, (c) and (d) $\varepsilon_u^*$,  position of the second peak.}
     \label{fit_second_peak}
 \end{wrapfigure}

\begin{figure}[h]
	\centering
	
	
	\includegraphics[width=0.7\linewidth]{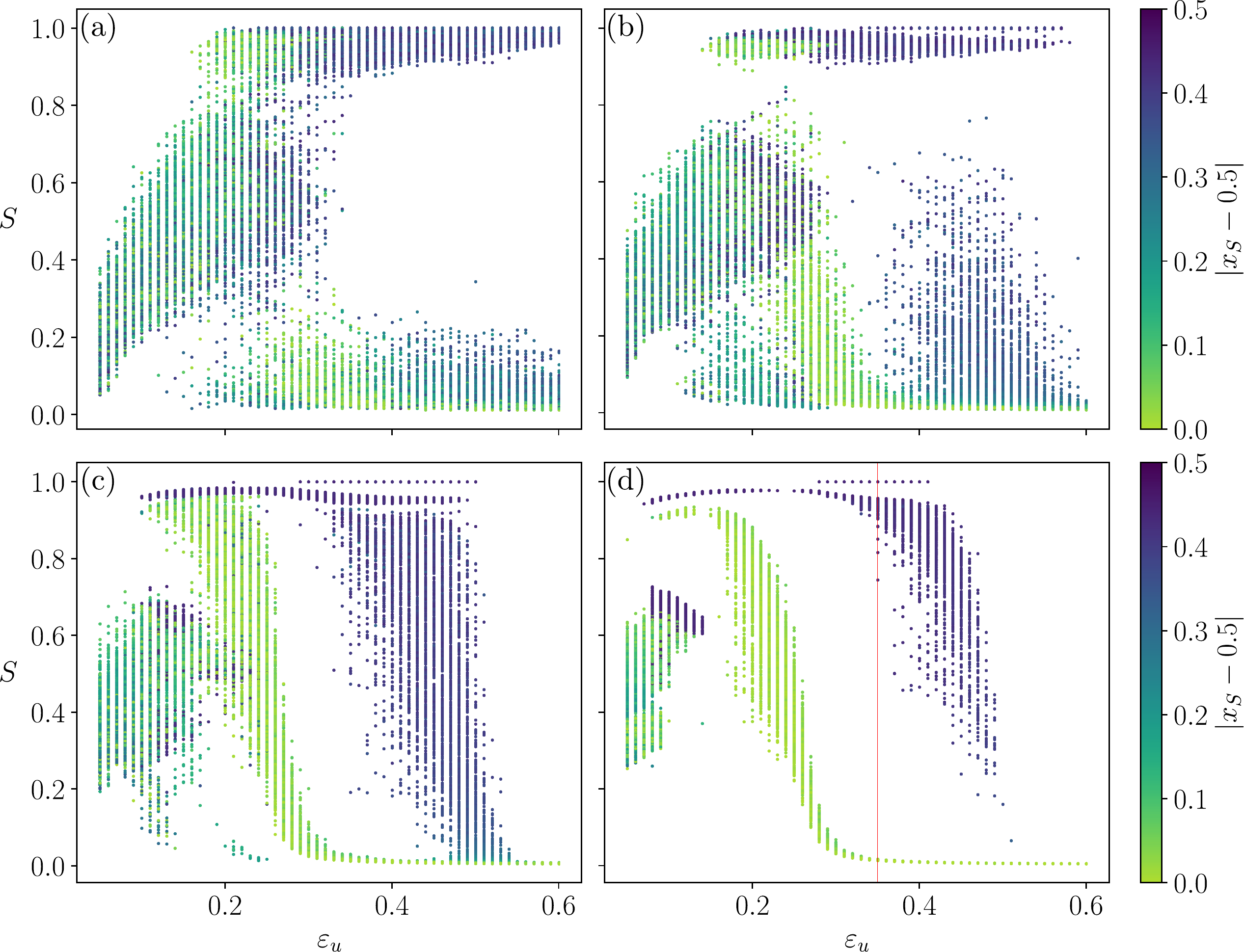}
	
	\captionsetup{justification=Justified}  
	\caption{Detail of size effects for the HK model in the Erdős–Rényi network of $\langle k \rangle=10$. Results for $1000$ samples, $\varepsilon_l=0.05$. (a) $N=256$, (b) $N=1024$, (c) $N=8192$, (d)  $N=92681$.}
	\label{scatter_ER_size_effect}
\end{figure}

In order to clarify this point it is necessary to  examine the distribution of largest clusters along all the realizations. In Fig.~\ref{scatter_ER_size_effect} we show,   for different system sizes, $N$, the  scatter plot of  the largest clusters' normalized sizes, $S$, for $1000$ realisations of the same society (same network, confidences drawn from the same  interval, $[\varepsilon_l=0.05,\varepsilon_u]$), in the steady state.
 The color code represents the \textit{extremism} of the  opinion, $e$, held by the members of the largest cluster of each realization, i.e. its shift from the central opinion,  which corresponds to  the steady state for consensus in the homogeneous system  ($x_S(t \longrightarrow \infty)=0.5$), where $x_S$ denotes the opinion of the members of the largest cluster.
 Finite size effects are very strong, changing qualitatively the  structure of the scatter plots which  stabilizes only at large  $N$,  revealing the presence of coexistent phases. Although finite size effects are well known in equilibrium thermodynamics, the results shown here,  point out  again at  the necessity of considering large enough sizes so that the observed properties  stabilize, which is unfortunately, still too often overlooked. Notice that in panel (b), corresponding to $N=1024$, a typical size studied in the literature,  the structure is very different from the stable one. 
Similar finite size effects are  observed for the other studied networks (see Supplementary Material sections II.A and II.B for videos of the evolution  the scatter plots due to finite size effect, for  ER and BA networks). \\

Focusing on the largest studied size, $N=92681$, shown in    Fig.~\ref{scatter_ER_size_effect} (d),  it is clear that the dispersion of  $S$ as well as the corresponding  opinions  strongly depend on the value of $\varepsilon_u$. In particular, for intermediate values of  $\varepsilon_u$, like the one marked by a red line in Fig.~\ref{scatter_ER_size_effect}(d), societies where the  largest cluster  converges to a mild opinion ($x_S (t \longrightarrow \infty) \approx 0.5$) coexist with others where the largest cluster holds an extremist opinion (either $x_S (t \longrightarrow\infty) \approx 0$ or $x_S (t \longrightarrow \infty) \approx 1$). We will see that the former do not form consensus at a given opinion but are constituted of a large number of agents holding \textit{different opinions} around the central one. This explains the very low values of  S in the scatter plot leading to  the collapse of the value of $\langle S \rangle$, observed in Fig.~\ref{sizes_effects}. This implies that  when the interaction among  heterogeneous agents is constrained by a network,  $\langle S \rangle$ is not a suitable order parameter.  When the network of possible interactions has the Barabási-Albert topology, the same general qualitative behaviour is observed. However, as mentioned above, the peak of the order parameter at 
 $\varepsilon_u^* \approx 0.35$,  which diminishes with the size in Erd\"os-Renyi networks Fig.~\ref{sizes_effects}(a),  stays for Barabási-Albert networks Fig.~\ref{sizes_effects}(b), as shown in Fig.~\ref{fit_second_peak}.\\

\begin{figure}[h]
	\centering
     \includegraphics[width=\linewidth]{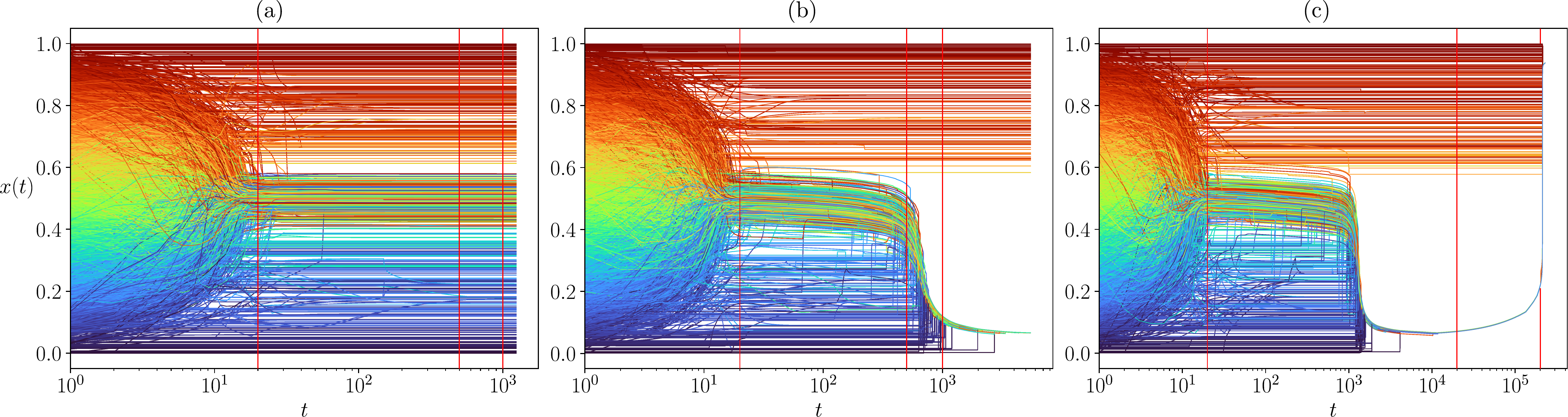}
     \captionsetup{justification=Justified}  
	\caption{Example of the opinion evolution of the agents  for a society where the interactions are constrained by an  Erdős–Rényi network of $\langle k \rangle = 10$, and with confidence interval $\varepsilon_l = 0.05$, $\varepsilon_u = 0.35$, $N=92681$. The vertical lines indicate the times corresponding to the snapshots shown in  Fig.~\ref{opinion_evolution_confidence} were taken. (a) Weak consensus, mild opinion phase: The agents located around the central opinion do not meet the criterion ($\delta x < 10^{-3}$) to constitute a single cluster. (b) Consensus extreme opinion phase. The final state is a dense strand containing the majority of the society while a few agents remain isolated. (c) Unanimous extreme opinion phase (U-turn): The opinion of the society is overturned from one extreme to the opposite one, the whole society is involved. Note that for readability and computational cost, only the evolution of 4000  agents taken at random  is displayed.}
	\label{trajectories}
\end{figure}

In order to understand the nature of the coexistent phases, we have  studied  individual trajectories corresponding to each of them.
Figure~\ref{trajectories} shows three examples of trajectories, one in each of the  observable phases at the point $(0.05,0.35)$ of the phase space (shown by the red line in  Fig.~\ref{scatter_ER_size_effect}(d)). These phases may be characterized by their opinion as :

\begin{itemize}
    \item \textit{weak consensus, mild opinion ("Mild")}: There is not a single large cluster of agents holding the same   opinion, instead a large number of agents hold \textit{neighbouring opinions} in the central region of the opinion space.
    \item \textit{consensus, extreme opinion ("Skewed')}: The opinion of the largest cluster lies far from  the central region.
    \item \textit{unanimity, extreme opinion ("U-turn")}: The opinion of the largest cluster lies at one of the extremes and the largest cluster includes almost all the agents in the population. 
   
\end{itemize}

In Fig.~\ref{trajectories}(a) we show the example of a  society that evolves towards a mild opinion. The final state is characterized by a broad strand where a large number of agents hold an opinion around $x=0.5$, however they are not close enough so as to form a unique cluster within the criteria used in this work to distinguish opinion differences (see Methods). This is the reason why the largest cluster size is low in the scatter plot, although the fraction of the population  holding  extremist opinions  is comparatively small. Nevertheless, these extremists prevent the convergence of the central agents  to a strong consensus.  

Figure~\ref{trajectories}(b) shows the example of a society which ends up in a skewed phase, showing a majoritarian extreme opinion. As the initial opinions are uniformly distributed in the interval, there is a symmetry around the central one and a tendency to this central opinion is visible at the beginning of the dynamics, however at some point, agents with a central opinion are attracted to either of the extremes, and the agents that are near the other extreme stop interacting  with the majoritarian strand. Nevertheless the extremist cluster contains a large majority of the society, hence the large values of $S$ with an extremist opinion in the scatter plot.

Of special interest is the formation of the U-turn  phase, shown in Figure~\ref{trajectories}(c). At an intermediate  stage, the  dynamics  is similar to that of the skewed phase; however  some  open minded agents in the majoritarian strand given their position in the network, are able to form a bridge between the  strand and the isolated extremist agents in the opposite side of the opinion interval. This small set of agents is enough to completely overturn the majority of the society into the opposed extremist opinion.

\begin{wrapfigure}[22]{l}{0.4\textwidth}
    \centering
	\includegraphics[width=\linewidth]{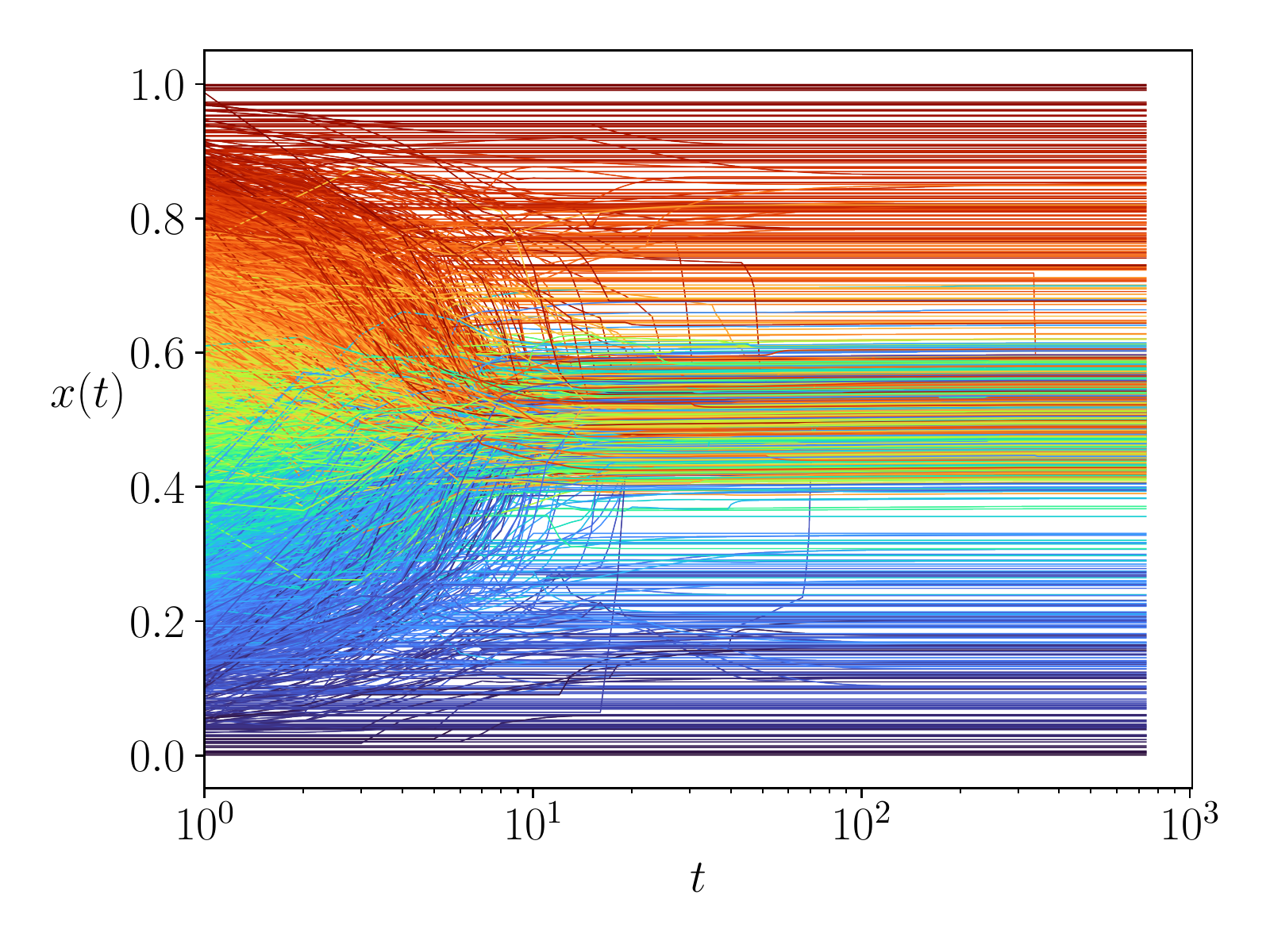}
    \captionsetup{justification=Justified}  
	\caption{Example of the opinion evolution of the agents  for a society where the interactions are constrained by an  Erdős–Rényi network of $\langle k \rangle = 10$, with a large proportion of open minded agents, $\varepsilon_l = 0.05$, $\varepsilon_u = 0.55$, $N=92681$.  Note that for readability and computational cost, only the evolution of 4000  agents taken at random  is displayed.}
	\label{trajectory_large_epsilon}
\end{wrapfigure}

Both extremist societies are the result of the interplay of the agents' heterogeneity with the network of potential interactions, and they are absent from both the heterogeneous mixed population case~\cite{schawe2020open} and the homogeneous model in networks~\cite{schawe2021bridges}. 
It should be noticed that the situation where the whole society may suddenly change from one extreme of the opinion space to the other, does not require the presence of  agents with very  large confidence values, the existence of a few agents with $\varepsilon_u \approx 0.3$ is enough for this to occur. This phenomenon is observed in both types of studied random networks (see Supplementary Material, section III.C, for examples corresponding to the Barabási-Albert network).

Let us now analyse  the behaviour of the system for large confidence values. At $\varepsilon_u >0.55$ the society contains many agents with large potential for compromising:  as they have large confidence values, they interact with many of their neighbours. It may seem paradoxical then,  that  all the realizations  have $S \longrightarrow 0$ in the scatter plot. Figure~\ref{trajectory_large_epsilon} shows that for large $\varepsilon_u$ the situation corresponds to the weak mild consensus: open minded agents rapidly integrate the central strand, which is pulled apart by the remaining closed minded ones that have been left aside on both extremes of the opinion interval. Again, this situation requires both agents heterogeneity and  the interactions being locally constrained by a network.\\

\begin{figure}[H]
	\centering
    \includegraphics[width=\linewidth]{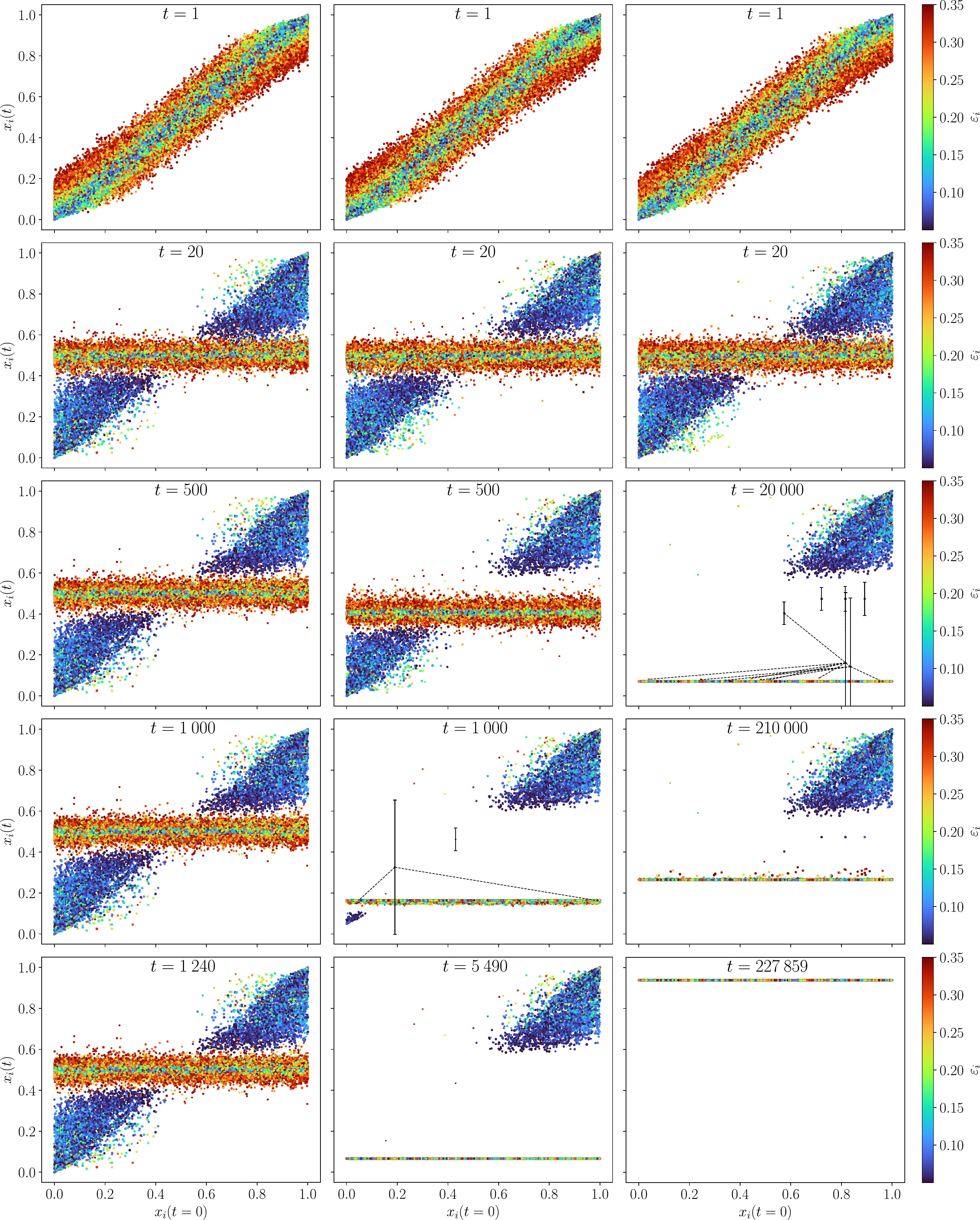}
    \captionsetup{justification=Justified}  
	\caption{ Snapshots of the evolution of the agents opinion as a function of their initial opinion, with the size of the dots proportional to the degree, color-coded by their confidence for a society of $N=92681$ agents where the interactions are constrained by an  Erdős–Rényi network of $\langle k \rangle = 10$, and with confidence interval $\varepsilon_l = 0.05$, $\varepsilon_u = 0.35$. Rows correspond to snapshots taken at different times. Notice that the  final convergence time is increasing from left to right.  Left column: Mild opinion phase.  Middle column: Skewed phase. Right column: U-turn phase. Bridges are highlighted in some frames: solid black vertical lines represent the confidence interval of the agents and dotted lines represent \textit{active links} (i.e. agents are topologically connected and within confidence reach).}
	\label{opinion_evolution_confidence}
\end{figure}

In order to illustrate how these phases are formed, Fig.~\ref{opinion_evolution_confidence} shows snapshots of the opinion of the agents at time $t$, $x_i(t)$, as a function their initial opinion, $x_i (0)$,  at the times  marked by vertical lines on the trajectories of Fig.~\ref{trajectories} ( see video in the Supplementary Material, section III.B and III.C for complete evolution).
The agents' confidences, $\varepsilon_i$ are also indicated by the color scale.

In the three cases the  central opinion region is quickly populated by agents with large confidences, while those with low confidences, and extreme initial opinions remain with extreme opinions for longer times (second row of Fig.~\ref{opinion_evolution_confidence}) . In the left column, corresponding to the  mild opinion phase, this situation remains stable and the central strand does not become dense because there are agents with extreme opinions and small confidences on both sides of it,  which prevent the central ones to derive towards either extreme. In the middle column, corresponding to the skewed phase,   we observe the same initial dynamics, but the agents in one extreme succeed in pulling down the whole group of mild opinion because many of the open minded agents within it are able to interact (and adopt) the extremist opinion while the extremist ones do not modify theirs. Finally  the right column shows the mechanism responsible for the U-turn trajectories which leads to an unanimous extremist phase. First, a large fraction of the society is pulled towards one extreme of the opinion interval, as in the previous case, but there are   open minded agents who can act as a bridge with the close-minded extremist ones in the opposite extreme of the opinion space. The latter finally  succeed in slowly pulling the whole society towards their opinion. The confidence interval of the open-minded agents who form the bridge are  represented by the vertical lines of Fig.~\ref{opinion_evolution_confidence}. These three dynamics have different time scales, the U-turn one being two order of magnitude longer that the others. 
The key ingredient that determines whether the dynamics will lead to a  skewed or to a  U-turn phase is the presence of \textit{active} bridges. At $t=1 000$ in the middle column of Fig.~\ref{opinion_evolution_confidence}, we can see the potential for a bridge: there is a very open-minded agent that is slightly "left behind"  from the main cluster and a close-minded agent that is within its confidence reach. However as there is no topological link between the two, the bridge cannot be "activated" and the phase remains skewed. On the contrary, at $t=20 000$ in the right column, the isolated open-minded agents and the close-minded  agents that act as  "anchor points", are topologically connected and within confidence reach. The bridges are therefore active, and through repeated interactions, they will end up pulling the entire main cluster toward the other opposite extremist opinion.

\begin{figure}[h]
	\centering
    \includegraphics[width=\linewidth]{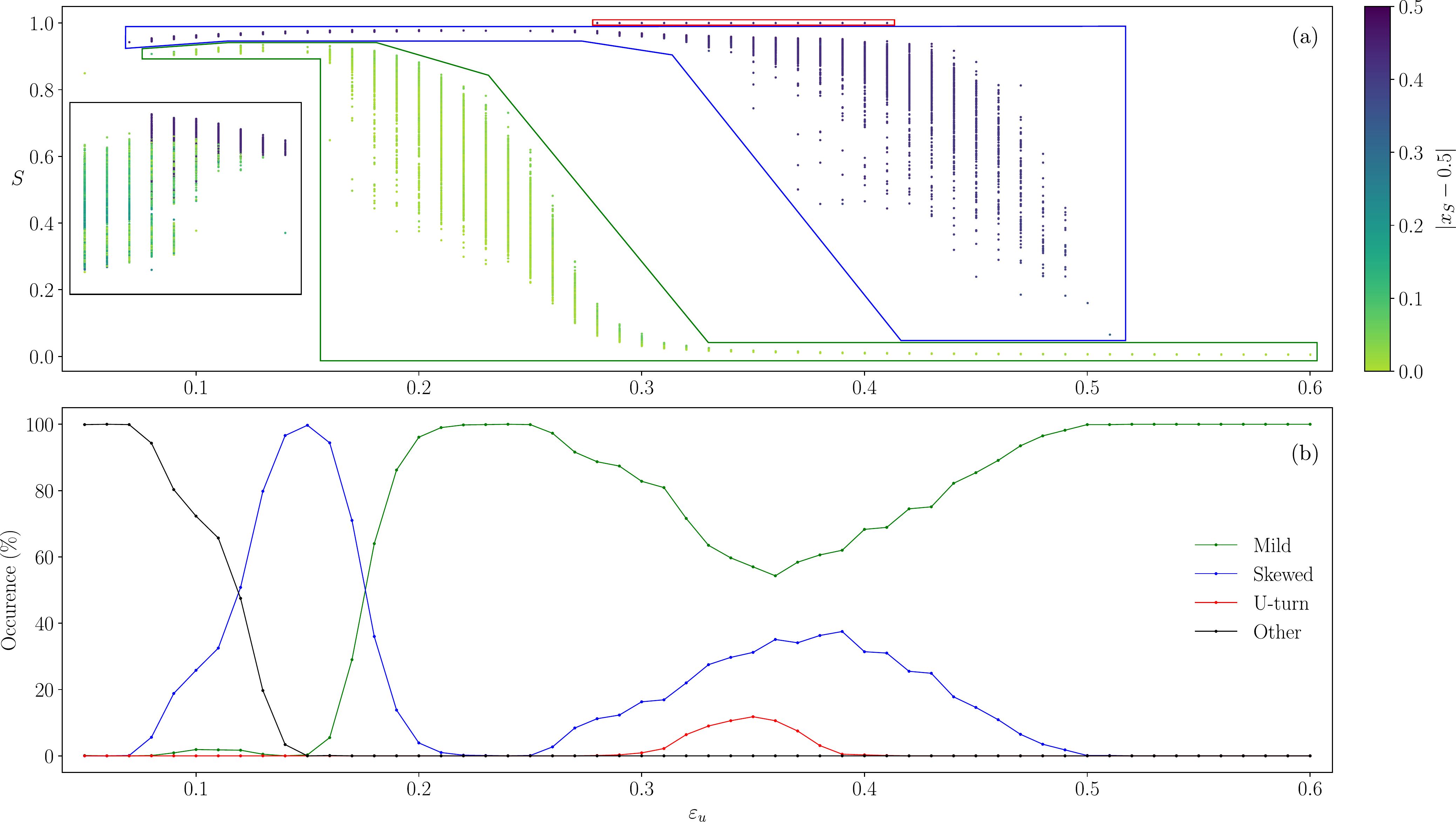}
    \captionsetup{justification=Justified}  
	\caption{Probability of each type of phase for societies having  $\varepsilon_l = 0.05$, as a function of $\varepsilon_u$, for $N=92681$ and  interactions  constrained by an Erdős–Rényi network of $\langle k \rangle = 10$. Averages calculated over 1000 samples. (a) The coloured boxes represent the samples used to compute the percentages shown in panel(b). (b) Probability of occurrence of each of the three phases observed in the steady state.  }
	\label{occurence_rate}
\end{figure}

Finally, although one cannot predict for a single sample which will be its steady state, it is possible to estimate the probability of occurrence of each of the three phases.  Figure~\ref{occurence_rate} shows that it  strongly depends on interval  $[\varepsilon_l$ , $\varepsilon_u]$ characterizing the society. 
The probabilities have been calculated by computing, for $\varepsilon_l=0.05$ and each $\varepsilon_u$, the fractions of  samples that end up in each of the three phases. To do so we consider a phase as "skewed" , when the opinion of the agents on the largest cluster is $\abs{x_S-0.5} > 0.3$.  The samples included in each case are shown by the coloured boxes of the scatter plot of Figure~\ref{occurence_rate}(a). The label "other" corresponds to the black box, where all agents of the society have very low confidences such that final state of the opinion remains in general, very fragmented.

In particular,  the probability of observing the society drastically changing from one extreme of the opinion space to the opposite one,  requires the presence of  agents who are  prone to compromise with others holding rather different opinions, although not with \textit{any} opinion. It should be noticed that the   values of $ $ at which  the U-turn phase occurs are smaller than the trivial value   $\varepsilon_u = 0.5$, which allows the interaction between the extreme opinions and the central one. In this region of the phase space the three phases may occur.

\begin{figure}[h]
	\centering
    \includegraphics[width=.75\linewidth]{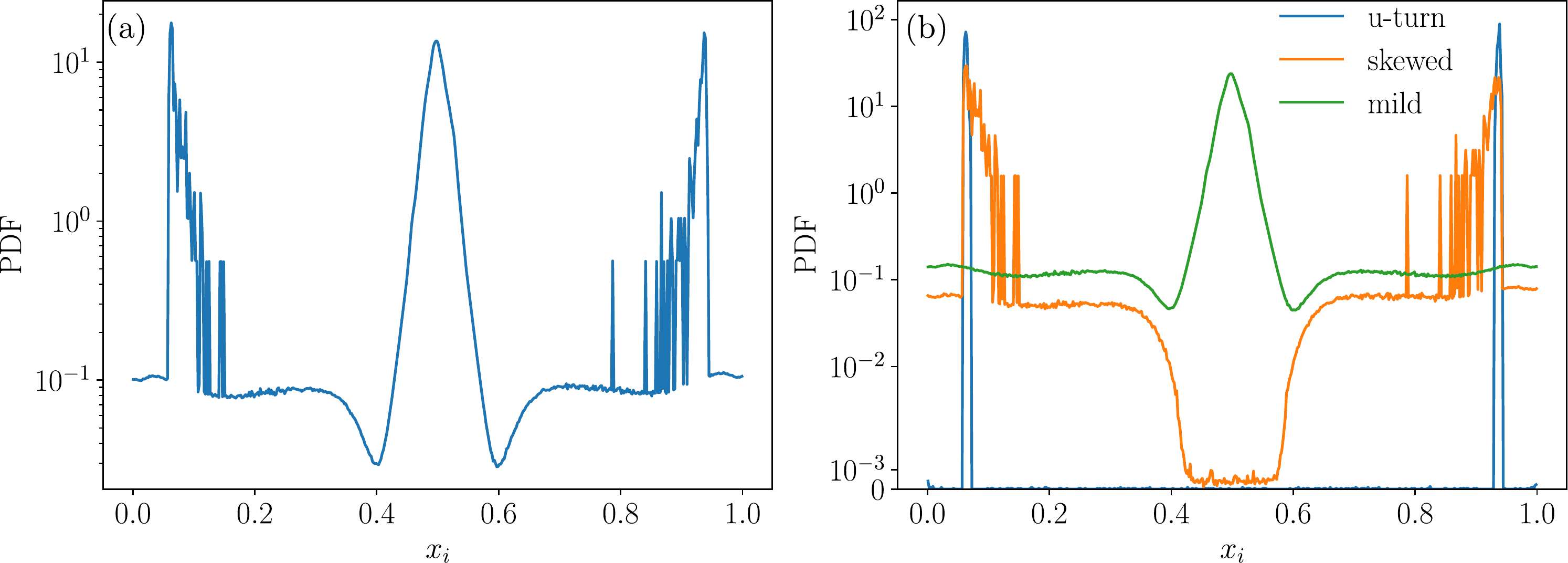}
    \captionsetup{justification=Justified}  
	\caption{Distribution of the final opinions of all the agents in a society of size $N=92681$, where the interactions are constrained by an  Erdős–Rényi network of $\langle k \rangle = 10$, and with confidence interval $\varepsilon_l = 0.05$, $\varepsilon_u = 0.35$, 1000 samples. (a) All realizations. (b) Realizations separated by the type of the final state: Weak consensus, mild opinion phase (mild); Consensus, extreme opinion phase (skewed); Uniformity, extreme opinion phase (u-turn).}
	\label{distrib_final_opinions}
\end{figure}

The distribution of the final  opinions for each of  these phases is shown in Figure~\ref{distrib_final_opinions} for the interval $[0.05,0.35]$. The distribution has been computed considering  the opinions of all the agents in the steady state (not only those corresponding to members of the largest cluster), for of all the realizations, thus  confirming the results illustrated by single trajectories of   Fig.~\ref{trajectories}. 

\subsection{Particular case: lattices}\label{cycle}
\begin{figure}[h]
	\centering
    \includegraphics[width=.5\linewidth]{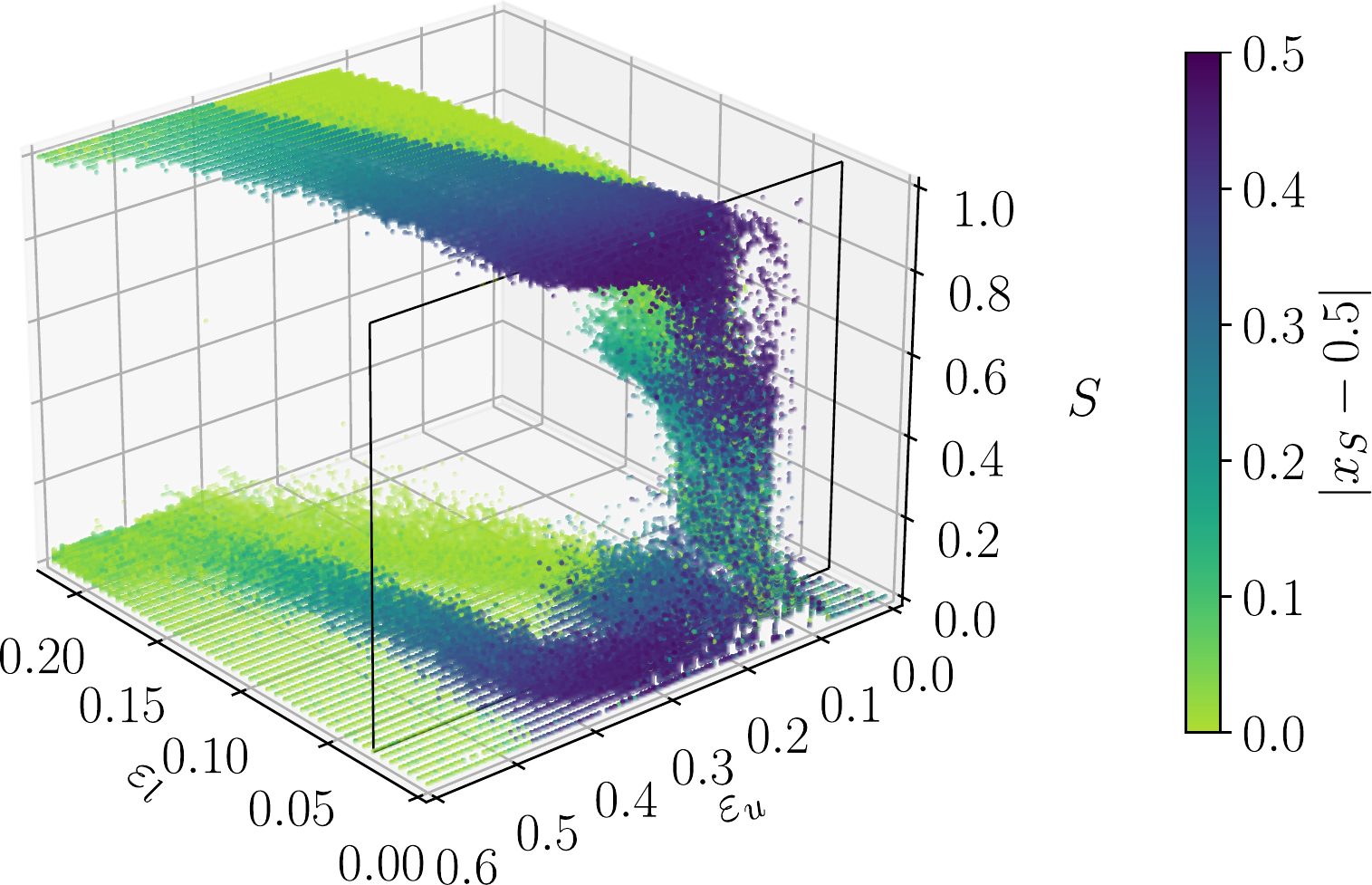}
    \captionsetup{justification=Justified}  
	\caption{3D representation of the scatter plot of the sizes of the largest cluster in the steady state for a system of size $N=16384$, in a Square Lattice with up to third neighbour interactions. The color map indicates the  extremism of the majoritarian cluster. The slice at $\varepsilon_l=0.03$ (vertical red line in  Fig.~\ref{S_comparo_topo}) is outlined in black.}
	\label{3D_Scatter_SL_k12}
\end{figure}

 For the sake of comparison,  we  have also studied the case  where the possible interactions are constrained by a regular lattice. The study of the homogeneous system has  shown that the outcomes of the HK dynamics are  qualitatively  different in lattices and in random networks~\cite{schawe2021bridges}. Figure~\ref{S_comparo_topo} shows that this remains true for the heterogeneous model. Fig.~\ref{3D_Scatter_SL_k12} shows the 3D scatter plot of the sizes of the largest clusters in the steady state,  for a system of $N=16384$, in a square lattice with up to third neighbours interactions, which leads to a coordination of $k = 12$, comparable to the average degree of the studied  networks.  
The main  difference with the latter is that, for lattices,  the coexistent 
weak and strong consensus phases contain agents whose opinions may be either mild or extremist, 
 while in networks we observe a clear separation of weak consensus around a mild opinion and strong consensus around extremist ones.

\begin{figure}[h]
    \centering
    \includegraphics[width=\linewidth]{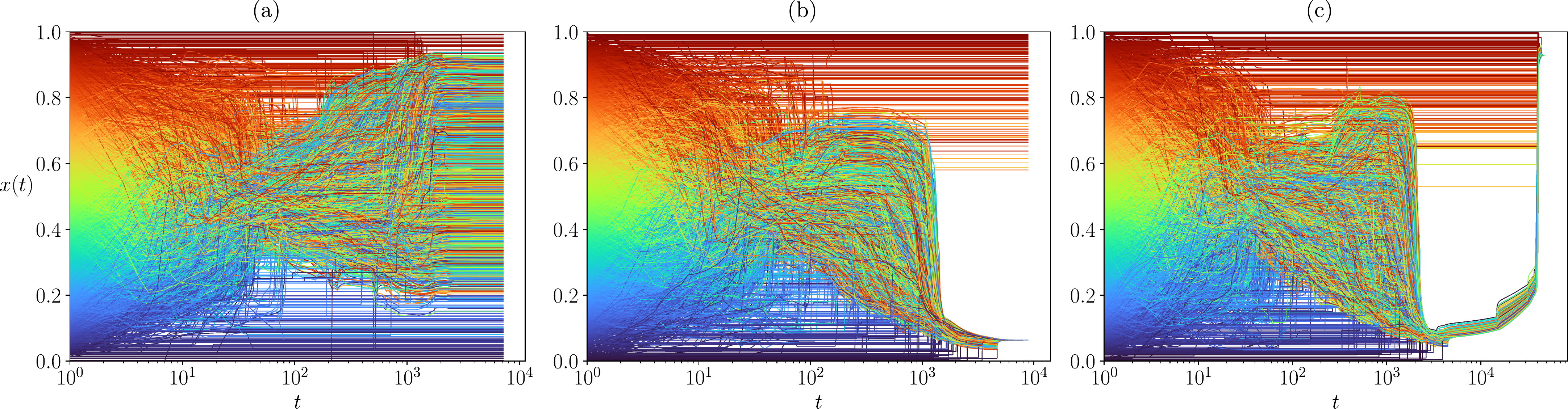}
    \captionsetup{justification=Justified}  
    \caption{Example of trajectories for a system of size $N = 16384$, in a Square Lattice with up to third neighbour interaction, with $\varepsilon_l = 0.03$, $\varepsilon_u = 0.35$. Note that for readability and computational cost, only the evolution of 4000  agents taken at random  is displayed. (a) Symmetric state around  final opinions  spanning a large central fraction of the opinion space, (b) Skewed extremist phase, (c) U-turn phase, notice the discrete jumps in opinion during the transient.}
	\label{trajectories_SL}
\end{figure}

The evolution of the agents is also very different from that of random networks, as shown by the trajectories of Fig.~\ref{trajectories_SL}, in particular because there is no initial  tendency to a (weak) consensus around a mild opinion. Here, the trajectories occupy  a very broad region of the opinion space.

Surprisingly, we have observed some "U-turn" trajectories. This is  counter-intuitive  because topological bridges do not exist in a lattice  (all agents are connected to the same amount of neighbours). In fact, we found that the mechanism leading to the massive opinion change  from  one extreme of the opinion space to the other is different from that of networks.  In networks, a few agents who have active links with two groups of opposite opinions, do not merge with any of them and succeed in bringing them together after a long transient. Here  we observe   successive jumps in opinion  of large domains of agents. It should be noticed that lattices are embedded in a physical space, giving rise to the notion of neighbours of different order (nearest neighbours, second order neighbours, etc). In order to have comparable average degree, the results presented here correspond to a square lattice with up to third neighbours interactions. The lattice contains spatial domains of different opinions that at a given time of the evolution are delimited by inactive links. However,  
agents inside one opinion domain may establish an active link  directly inside the other opinion domain, via second or third neighbours  interactions, allowing for a further evolution that might overturn the global opinion state from one extreme to the other (see videos in section III.D of the Supplementary Material). 

A particularity of lattices is the existence of some rare samples in the region of very low $\varepsilon_l$ which do not converge. In these samples, we observe  a small, localized region of the lattice that  enters in a  cycle where the same group of neighbours keeps varying their opinion periodically.  The  confidence interval where such samples are found, is represented by a white pixel in the phase diagram of Fig.~\ref{S_comparo_topo} to indicate that, due to this effect, we have not computed the average size of the largest cluster, $\langle S \rangle$. Randomness in the connections destroys this effect.  3D animations of the phase diagrams coded by the extremism of the final opinion,  for all studied topologies, can be found in the Supplementary Material for a detailed comparison. 

In order to compare with the random networks, we  explored in detail different societies with $\varepsilon_l=0.03$ (vertical red line in  Fig.~\ref{S_comparo_topo}), a region which shows similar $\langle S \rangle$ behaviour as the random configurations as $\varepsilon_u$ increases. In this region typical trajectories are shown in Fig.~\ref{trajectories_SL}.
 
\section{Discussion}\label{discu}

In this work we study a bounded confidence model of  opinion dynamics, which combines heterogeneity in the agents' idiosyncratic properties, materialized by a quenched disorder, with interactions that are constrained by a network of social contacts. 
Our results show that, in presence of  very closed minded agents, the existence of others that have a tendency to   compromise with their neighbours holding an  opinion quite  different from theirs, may lead a large amount of the  population that started to move to an  intermediate opinion, to derive towards an extremist one. More spectacularly, if some of the compromising agents find among its social contacts, another one which is in turn interacting   with others holding a minoritarian opinion on the opposite extreme, the whole population may be overturned from one extreme to the other.

This result is relevant for applications in real settings which involve randomness in social ties and heterogeneity in the agents' properties. It calls the attention about the usage of popular concepts as "compromise" and "open mindedness", which are usually endowed with a positive connotation, as they are considered  leading properties which  decrease conflict and promote consensus in society, avoiding extreme positions. Here we show that this  tendency to compromise with a quite different opinion may be paradoxically profitable to closed minded extremists. It is important to notice that this phenomenon occurs for societies with agents holding confidence values that are well below the trivial value of $\varepsilon = 0.5$, which would allow even the most extremist agents to interact with the middle opinion.

This phenomenon that results from the  interplay between the heterogeneity the agents' confidences  and in their connections, is  absent from both  the heterogeneous model in the mixed population and the homogeneous model in networks. The key ingredient being the possibility to bridge among two different opinion clusters, where one of them contains open minded agents and the other  closed minded ones. The complete overturn of the society from an opinion to another requires longer time scale. This is an issue of practical importance if one considers that, in real systems, the evolution time may be limited in such a way that the system can or cannot reach a steady state. This is typically the case of the time let to the population to form an opinion during electoral campaigns.

This article  also  questions the pertinence of a standard order parameter, $\langle S \rangle$, largely used in the study of opinion dynamics models~\cite{Castellano2009Statistical}. It also recalls  the importance of the size effects, which unfortunately are not sufficiently taken into account in the literature, where typical population  sizes are around $N=1000$ and rarely go over $N=10000$. Most of  the findings shown here will have remained hidden had we limited our study to the behaviour of the standard order parameter on these typical sizes.

Our results allow us to  statistically characterize the probability of observing a trajectory leading to a weak consensus around the mild opinion, to a strong  consensus around one extreme opinion, or to an U-turn trajectory which overturns the society's majoritarian opinion from one extreme to an unanimous agreement about the opposite extreme, according to the value of the confidences. Nevertheless, the question remains as to whether it would be possible to  predict the final state
from the properties of the agents, for instance, their confidence and position on the network.  Unsurprisingly, a first machine learning test using a simple multi-layer perceptron architecture,  did not lead to conclusive results, given that the number of total realisations ($1000$) is good for the statistical purposes, but remains small to train and test a machine learning algorithm. 
One important  point is that  the bridges, which here involve not only the position in the network and the confidence of the agent creating the bridge, but also its dynamical opinion value,  occur during the evolution, therefore  the details of the dynamics, and not only the quenched variables and initial conditions,  intervene to create "U-turn" trajectories. 

A next step along this line would be to introduce correlations between the agents' properties and  their social position, measured by their degree or their centrality. Would it then be possible to "tailor" the system so as to be able to drive the society's opinion to a given region of the opinion space? This  work is in progress.

\section{Conclusions}
In the road to  more realistic models of opinion dynamics, heterogeneity is an unavoidable issue. We see here how the simplest possible combination of heterogeneity -in the agents' properties and in their interactions- leads to novel, almost paradoxical results. These results also raise the question about the attributes usually associated to individual attitudes:  the tendency to compromise, abandoning one's ideas or opinions is usually considered positively, arguing that it allows  a soft evolution to a consensus that will prevent the society from undergoing strong and radical changes. This works shows that this is not always the case, and that the final result  depends on where  the open and closed minded agents are placed in the network. This point is particularly relevant when one takes into account that that nowadays, a large amount of the population builds its opinion on different subjects based on information circulating through different social networks. The links on the  platforms giving access to these social networks are mediated by the algorithms which control to what information agents are exposed, coupling the heterogeneity of the connections with the dynamics of the opinion.

\section{Model \& Methods}\label{sec1}
\subsection{Hegselmann-Krause model}\label{HKmodel}
We study the Hegselmann-Krause model (HK), which describes a compromise dynamic under bounded confidence. Each   agent $i$, $i = 1, N$, of the population is endowed of a dynamical continuous variable $x_i(t)\in[0,1]$ representing its \textit{opinion}. In its heterogeneous version~\cite{Lorenz2010Heterogeneous,schawe2020open} each agent is characterized by a quenched variable $\varepsilon_i$, modeling the \textit{confidence} of the agent, i.e. its aptitude to interact with others holding different opinions. Every agent $i$ may interact with any other agent $j$ provided that their opinion differs
in less than the confidence range $\varepsilon_i$, i.e. $x_j \in [x_i - \varepsilon_i, x_i + \varepsilon_i]$. 

Here, we study the HK model, additionally constrained by the topology of an underlying static network that materializes the possible social contacts of the agents.
Formally, we define the model on a graph $G = (V, E )$, where $V$ is a set of vertices and $E$ a set of non-directed edges, representing the agents and their  interaction's possibilities, respectively. The set of interaction partners of agent $i$, its \textit{neighbourhood}, taking into account both the network constraint and the  confidence $\varepsilon_i$, is therefore:
\begin{equation}
I(i, \overrightarrow{x} ) = \{1\leq j \leq N | \lvert x_i - x_j \rvert \leq \varepsilon_i \wedge \{i,j\}\in E\}.
\label{Eq1}
\end{equation}
 The dynamics is defined in discrete time by a synchronous update of all agents at each time step. In each time step an agent $i$ adopts the average opinion of all neighbours in the set defined by (\ref{Eq1}), i.e.
\begin{equation}
x_i(t+1) = \frac{1}{\lvert I(i, \overrightarrow{x}(t))\vert} \sum_{j\in I(i, \overrightarrow{x}(t))} x_j(t).
\label{Eq2}
\end{equation}

The networks of social ties are required to be connected simple graphs, i.e., no multiedges or self-loops. Notice that, as in the original HK model, the active agent is explicitly included in the average of (\ref{Eq2}), which is required to prevent the synchronous update to induce artificial limit cycles in the dynamics.

For each run we start by drawing the initial opinions and the confidences from uniform distributions: $x_i(t=0) \in \mathcal{U}[0,1]$ and $\varepsilon_i \in \mathcal{U}[\varepsilon_l, \varepsilon_u]$ with $\varepsilon_l \leq \varepsilon_u$, $\forall i=1,...,N$ and both $\varepsilon_l, \varepsilon_u \in [0,1]$.

The simulations are run by iterating (\ref{Eq2}) until the opinions (represented with 64-bits IEEE 754 floating-point numbers) converge, reaching a final \textit{steady state}. The convergence criterion we use here requires that the sum of the changes over all agents is below a threshold, i.e
\begin{equation}
\sum_{i=1}^{N} \abs{x_i(t-1) - x_i(t) } < 10^{-4}.
\end{equation}

The main observable of interest is the relative size $S$ of the largest cluster, a cluster being defined as the set of agents holding the same opinion within a tolerance of $10^{-4}$.

Unless stated otherwise, in the exploration of the full parameter space at each point $(\varepsilon_l, \varepsilon_u)$, results computed by  averaging over 100 realizations, while for the finite-size analysis at a given $\varepsilon_l$,  averages are taken over 1000 independent realizations of the same society (same network type, and confidences $\varepsilon_i$ drawn from the same   $[\varepsilon_l, \varepsilon_u]$ interval), each of which is called a \textit{sample}.

Along with the average relative size of the largest cluster $\langle S \rangle$, we also compute its variance $\mathrm{Var}\left(S\right)$ and the average convergence time $\langle T \rangle$, we also defined the \textit{largest cluster size} entropy  as: 
\begin{equation}
    \sigma = -\sum_{z\in Z} p_z \ln(p_z)
\end{equation}
where $p_z$ is the discrete probability for a given realization to present a largest cluster of size $z$, normalized to $\sum_{z\in Z}p_z = 1$, and $Z$ is the set of all occurring largest cluster sizes across all the considered realizations. Note that this differs from the usual cluster size entropy, which would be defined with $p_s$ the probability for an arbitrary agent to be member of a cluster of relative size $s$. Our definition of the entropy will differentiate  regions of the parameter space where the final state may present a broad or a sharp distribution of sizes, $S$, of the largest cluster of the different samples.

We introduce another metric, the \textit{extremism}, $e$, defined as:
\begin{equation}
    e = \abs{ x_S - 0.5 }
\end{equation}
where $x_S$ is the opinion held by the cluster of largest relative size $S$. As the opinion space is agnostic and symmetric with respect to the value $x=0.5$, we subtract 0.5 and take the absolute value. The extremism is thus defined $e \in[0,0.5]$ where $e=0$  denotes the majoritarian group having a totally \textit{mild} opinion $x_S=0.5$, while $e=0.5$ means that the majoritarian group has an opinion at one extreme or the other of the opinion space:  $x_S=0$ or $x_S=1$.

\subsection{Studied networks}\label{networks}
We study a selection of network models chosen to cover different topological properties going from lattices to random networks, and from fully connected to scale-free networks. The  networks considered here are~\cite{newman_networks}:\\

\paragraph{Fully connected network}\label{fc}
Agents arranged in a complete graph can be considered as not being restrained by any topology at all, as each agent is able, in principle, to interact with any other agent in the society. This is why it is also referred to as the mixed population society. This corresponds to the setup of the original HK model.\\

\paragraph{Square lattice (SL)}\label{sl}
We use periodic boundary conditions so as to ensure perfect regularity of the neighborhood of the agents. We vary the connectivity from nearest to third neigbours interactions, the latter giving $\langle k \rangle = k = 12$ which can be compared to the studied random networks with $\langle k \rangle =10$. At a difference with other network topologies, lattices are embedded in a $D$-dimensional space. Here we concentrate our study on $D = 2$. Lattices show other important differences with respect to random networks: they have a long average path length and a large clustering coefficient.\\

\paragraph{Barab\'{a}si-Albert (BA)}\label{ba}
This network model connects vertices using a preferential attachment procedure and leads to a scale-free degree distribution with a slope of $-3$ in the large-$N$ limit . This growing network model starts with a clique of $m$ nodes and adds sequentially new agents, each of which brings $m$ new edges. Each of these edges is connected to the existing core with a probability proportional to the current degree of the target node. In this way nodes with a high degree gain more neighbors in a “the rich get richer” procedure. By construction, the networks of this ensemble have mean degree $\langle k\rangle = 2m$.\\

\paragraph{Erd\H{o}s-R\'{e}nyi (ER)}\label{er}
Also known as uniform or binomial random graph, any two nodes are connected with probability $p$, leading to a network with uncorrelated connections and a mean degree $\langle k \rangle=Np$ , in the large-$N$ limit. Here we study the sparse version for a fixed value of $N p$ . For large $N$ one needs to control that the generated networks are connected in order to avoid misleading results of the HK dynamics. Since we are conditioning on connectedness, the generation of ER realizations may not be trivial. Especially, it is known that almost all ER realizations are not connected if $N p  < \ln(N)$, in the limit of large $N$. In our case, we only study $N p  = 10$, which prohibits a rejection based sampling of connected ER realizations at $N \geq 16384$. However, the results for sizes up to $ N = 92681$ [which is well above the threshold as $\ln(92681) \approx 11.4$, but apparently close enough that we encounter enough realizations of connected ER] already draw a convincing picture, such that we do not need to use more sophisticated methods, like Markov chain methods, to generate connected ER of larger sizes.

\section*{  \textbf{Data  Availability} }
	The datasets generated during and/or analysed during the current study are available at:  \href{https://doi.org/10.5281/zenodo.7455640}{doi.org/10.5281/zenodo.7455640}.
	
	

	
	\section*{ \textbf{Author Contributions} }
	
	R.P.: created and executed the code, designed 3D visualizations and animations, analysed and discussed  the results. L.H.: conceived the research questions, designed the research protocol,  discussed  the results and wrote the manuscript. H.S.: discussed results and provided feedbacks during the initial conception of the code, ALL: revised the manuscript. \\
	
	\section* {\textbf{Competing interests}} 
	
	The authors declare no competing interests.\\
	
	\section*{Acknowledgments}
	The authors acknowledge the OpLaDyn grant obtained in the 4th round
	of the Trans-Atlantic Platform Digging into Data Challenge (2016-147 ANR OPLADYN TAP-DD2016)
	and Labex MME-DII (Grant No. ANR reference 11-LABEX-0023).

\bibliography{lit}
\end{document}


\title{Supplementary Material  of "On the side-effects of compromising: coupling agents' heterogeneity with network effects on a bounded confidence opinion dynamics model."}
\author{Rémi Perrier}
\email{remi.perrier@cyu.fr}
\affiliation{Laboratoire de Physique Th\'{e}orique et Mod\'{e}lisation, UMR-8089 CNRS, CY Cergy Paris Universit\'{e}, France}
\author{Hendrik Schawe}
\email{hendrik.schawe@cyu.fr}
\affiliation{Laboratoire de Physique Th\'{e}orique et Mod\'{e}lisation, UMR-8089 CNRS, CY Cergy Paris Universit\'{e}, France}
\author{Laura Hern\'{a}ndez}
\email{laura.hernandez@cyu.fr}
\affiliation{Laboratoire de Physique Th\'{e}orique et Mod\'{e}lisation, UMR-8089 CNRS, CY Cergy Paris Universit\'{e}, France}

\date{\today}
\maketitle
Due to a known issue with arXiv, hyperlinks are not available. All videos mentioned in this supplementary material are available at: \href{https://doi.org/10.5281/zenodo.7455640}{doi.org/10.5281/zenodo.7455640}
\section{Phase plots}
\subsection{3D visualisation}

In this section we present  videos that allow to inspect in 3D the scatter plots leading to the phase diagrams of Fig.1 of the main text, for different topologies. Each dot represents the normalized size of the largest opinion cluster at the steady state (on the Z-axis) for a given realization for a society where the confidences of the agents are drawn with uniform probability in the interval $[\varepsilon_l, \varepsilon_u]$ (X-axis and Y-axis respectively). The color code represents the level of extremism of the opinion of the largest cluster, $x_S$, given by $|x_S-0.5|$ (green: $|x_S-0.5|=0$, blue: $|x_S-0.5|=0.5$).

\begin{itemize}
    \item 360 rotation of 3D scatter plot for for a mixed population (fully connected network) space, $N=16384$.\\
    \href{https://zenodo.org/record/7455641/files/scatter3D_el_eu_Smax_uni_FC_N%3D16384.mp4?download=1}{\texttt{scatter3D\_el\_eu\_Smax\_FC\_N=16384.mp4}}\\
    
    \item 360 rotation of 3D scatter plot for a society constrained by an Erd\"{o}s Rényi (ER) topology with $\langle k \rangle$=10, $N=16384$.\\
    \href{https://zenodo.org/record/7455641/files/scatter3D_el_eu_Smax_uni_ER_k%3D10_N%3D16384.mp4?download=1}{\texttt{scatter3D\_el\_eu\_Smax\_ER\_k=10\_N=16384.mp4}}\\
    
    \item 360 rotation of 3D scatter plot for a society constrained by a Barabasi-Albert (BA) topology with $\langle k \rangle$=10, $N=16384$.\\
    \href{https://zenodo.org/record/7455641/files/scatter3D_el_eu_Smax_uni_BA_k%3D10_N%3D16384.mp4?download=1}{\texttt{scatter3D\_el\_eu\_Smax\_BA\_k=10\_N=16384.mp4}}\\
   
    \item 360 rotation of 3D scatter plot for a society constrained by a Square Lattice (SL) topology with $\langle k \rangle = k$=4, 8, 12, $N=16384$.\\
    \href{https://zenodo.org/record/7455641/files/scatter3D_el_eu_Smax_uni_SL_k%3D4_N%3D16384.mp4?download=1}{\texttt{scatter3D\_el\_eu\_Smax\_SL\_k=4\_N=16384.mp4}}\\
    \href{https://zenodo.org/record/7455641/files/scatter3D_el_eu_Smax_uni_SL_k%3D8_N%3D16384.mp4?download=1}{\texttt{scatter3D\_el\_eu\_Smax\_SL\_k=8\_N=16384.mp4}}\\
    \href{https://zenodo.org/record/7455641/files/scatter3D_el_eu_Smax_uni_SL_k%3D12_N%3D16384.mp4?download=1}{\texttt{scatter3D\_el\_eu\_Smax\_SL\_k=12\_N=16384.mp4}}\\
\end{itemize}

\subsection{Lattices with different coordination number} 

Here we present, as a complement of Fig.1 of the main text, the phase diagrams for the square lattice topology where the interactions extend to  first, second and third neighbours. 

\begin{figure}[H]
	\centering
	\includegraphics[width=\linewidth]{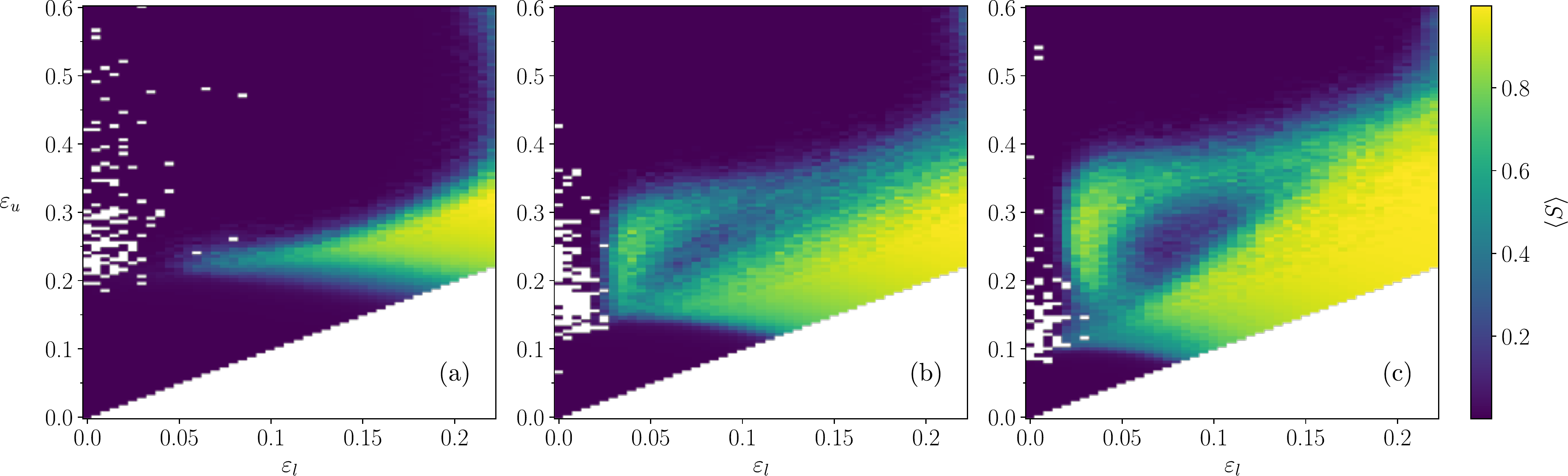}
	\caption{Phase plots for a society where the interactions are limited by a square lattice, for a system with $N=16384$ agents. The color map  gives the average relative size of the largest cluster $\langle S \rangle$ averaged over 100 realizations.
	(a)Interactions limited to nearest neighbours in the lattice,  $\langle k \rangle = k = 4$, (b)Interactions up to next nearest neighbours in the lattice, $\langle k \rangle = k = 8$, (c)Interactions up to third neighbours in the lattice, $\langle k \rangle = k = 12$. Note that some realizations did not converge in reasonable computing time according to the convergence criterion described in the Methods section of the main text. In order to avoid selection bias in the averaging, the  corresponding points of the phase space are marked in white.}
	\label{comparo_SL_phase_plot}
\end{figure}

\section{Study of finite size effects}
\subsection{3D visualisation as a function of $N$}

In this section we present videos that allow to inspect the scatter plots leading to Fig.2 of the main text, revealing the finite size effects,  for different topologies in 3D. The Z-axis and the colormap are the same, representing the extremism.

\begin{itemize}
    \item ER network with $\langle k \rangle$=10, $\varepsilon_l=0.05$\\
    \href{https://zenodo.org/record/7455641/files/scatter3D_el%3D0.05_eu_Smax_extremism_ER_k%3D10_SizeEffect.mp4?download=1}{\texttt{scatter3D\_el=0.05\_eu\_Smax\_extremism\_ER\_k=10\_SizeEffect.mp4}}\\
    
    \item BA network with $\langle k \rangle$=10, $\varepsilon_l=0.05$\\
    \href{https://zenodo.org/record/7455641/files/scatter3D_el%3D0.05_eu_Smax_extremism_BA_k%3D10_SizeEffect.mp4?download=1}{\texttt{scatter3D\_el=0.05\_eu\_Smax\_extremism\_BA\_k=10\_SizeEffect.mp4}}\\
    
\end{itemize}

\subsection{2D visualisation as a function of $N$}

This section presents the same results as the previous section, but here visualized in 2D, for the values of confidence marked by a vertical line on Fig. 1 of the main text. Panels of Fig.4 in the main text are essentially still-frames from such a video.

\begin{itemize}
    \item ER network with $\langle k \rangle$=10, $\varepsilon_l=0.05$.\\
    \href{https://zenodo.org/record/7455641/files/scatter2D_el%3D0.05_eu_Smax_extremism_ER_k%3D10_SizeEffect.mp4?download=1}{\texttt{scatter2D\_el=0.05\_eu\_Smax\_extremism\_ER\_k=10\_SizeEffect.mp4}}\\
    
    \item BA network with $\langle k \rangle$=10,, $\varepsilon_l=0.05$.\\
    \href{https://zenodo.org/record/7455641/files/scatter2D_el%3D0.05_eu_Smax_extremism_BA_k%3D10_SizeEffect.mp4?download=1}{\texttt{scatter2D\_el=0.05\_eu\_Smax\_extremism\_BA\_k=10\_SizeEffect.mp4}}\\
    
\end{itemize}

\subsection{Other metrics}

This section contains the results presented in Fig.2 along with additional metrics, namely: the average converging time according to the criterion explained in the section Methods of the main text; the variance of $<S>$; and the entropy of the size distributions  of the largest clusters, S.

\begin{figure}[h]
	\centering
	\includegraphics[width=\linewidth]{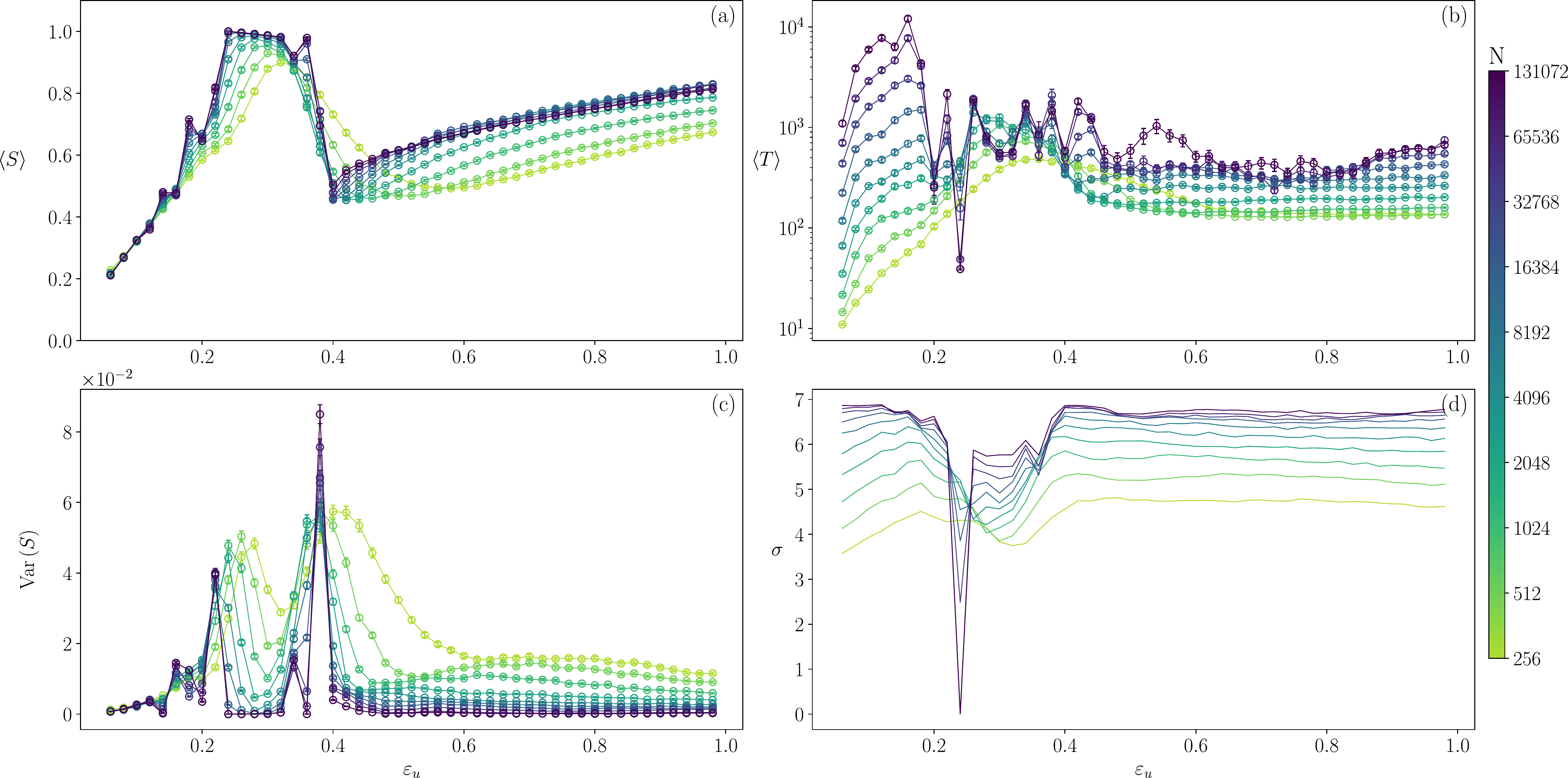}
	\caption{Metrics for the fully connected network with $\langle k \rangle=10$, $\varepsilon_l=0.05$. Average over 1000 realizations. Data taken with the permission of the authors from~\cite{schawe2020open}.
	(a) Average relative size of the largest cluster $\langle S \rangle$, (b) variance $\mathrm{Var}\left(S\right)$, (c) average convergence time $\langle T \rangle$, (d) largest cluster size entropy $\sigma$.}
	\label{metrics_FC}
\end{figure}

\begin{figure}[H]
	\centering
	\includegraphics[width=\linewidth]{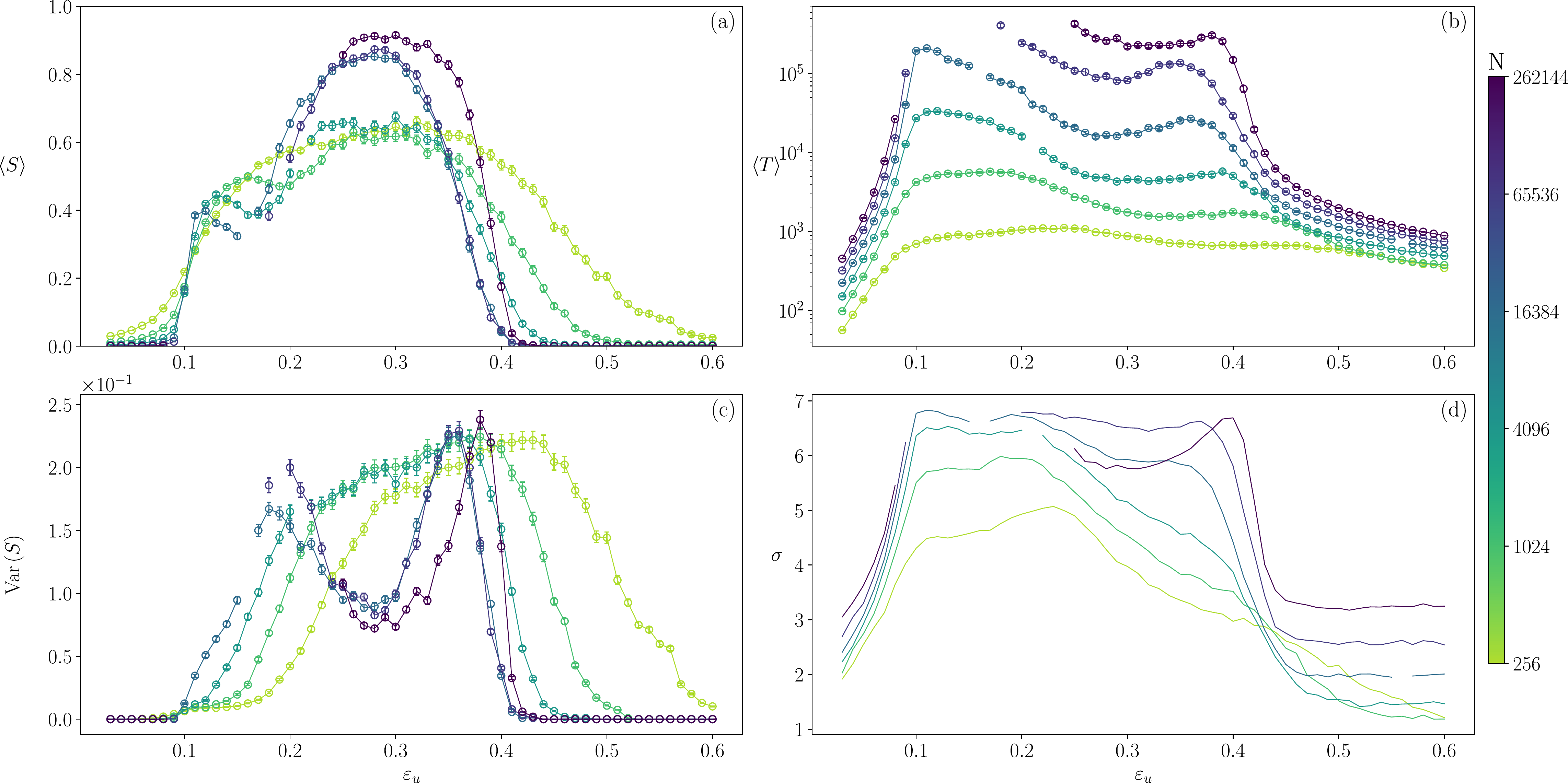}
	\caption{Metrics for the square lattice including up to third neighbour interactions ($k = 12$), $\varepsilon_l=0.03$. Average over 1000 realizations. 
	(a) Average relative size of the largest cluster $\langle S \rangle$, (b) variance $\mathrm{Var}\left(S\right)$, (c) average convergence time $\langle T \rangle$, (d) largest cluster size entropy $\sigma$. Note that some realizations did not converge in reasonable computing time according to the convergence criterion described in the Methods section of the main text. In order to avoid selection bias in the averaging, the  corresponding points are omitted.}
	\label{metrics_SL}
\end{figure}

\begin{figure}[H]
	\centering
	\includegraphics[width=\linewidth]{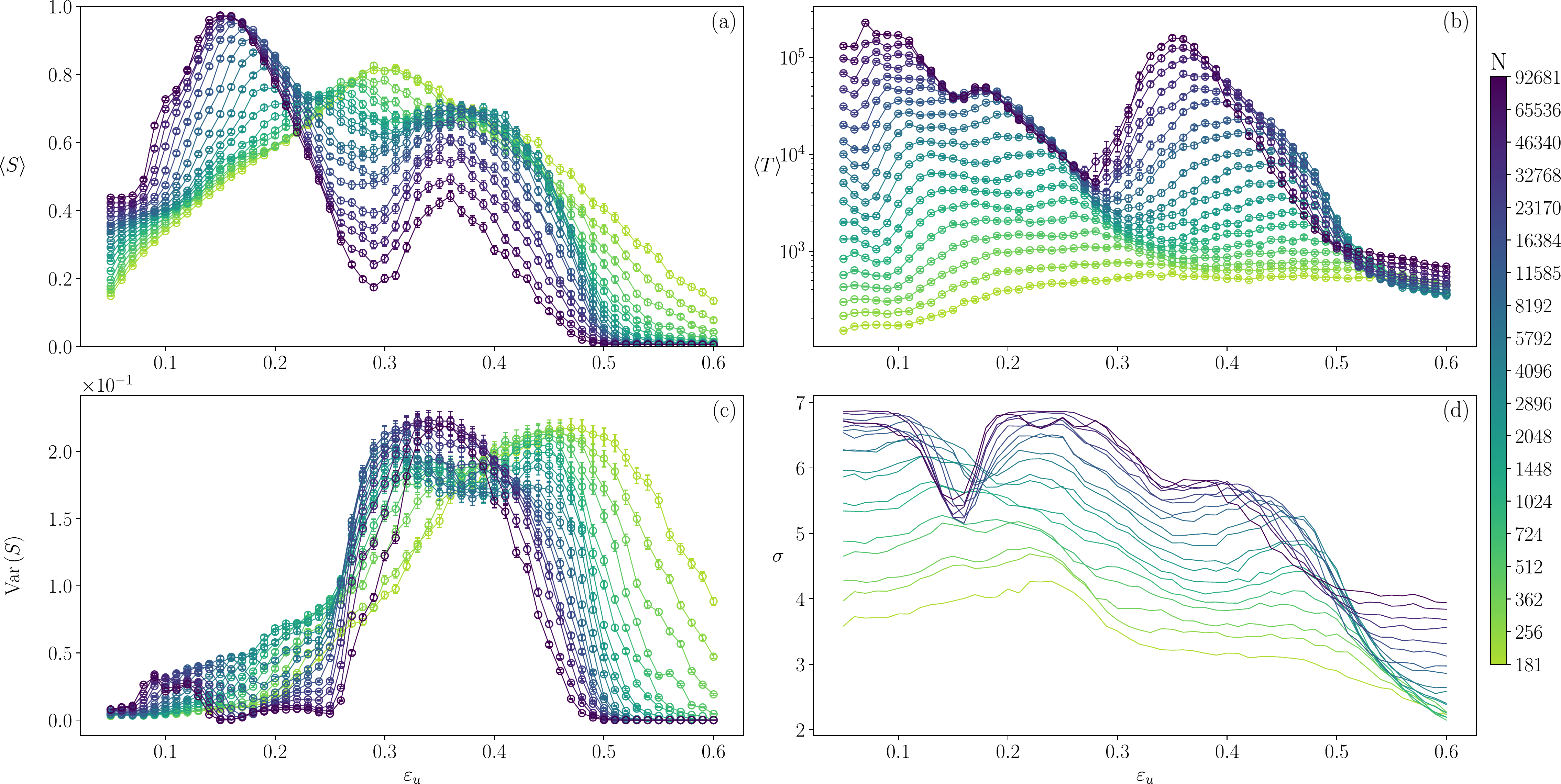}
	\caption{Metrics for the Erdős–Rényi network with $\langle k \rangle=10$, $\varepsilon_l=0.05$. Average over 1000 realizations.
	(a) Average relative size of the largest cluster $\langle S \rangle$, (b) variance $\mathrm{Var}\left(S\right)$, (c) average convergence time $\langle T \rangle$, (d) largest cluster size entropy $\sigma$. }
	\label{metrics_ER}
\end{figure}

\begin{figure}[h]
	\centering
	\includegraphics[width=\linewidth]{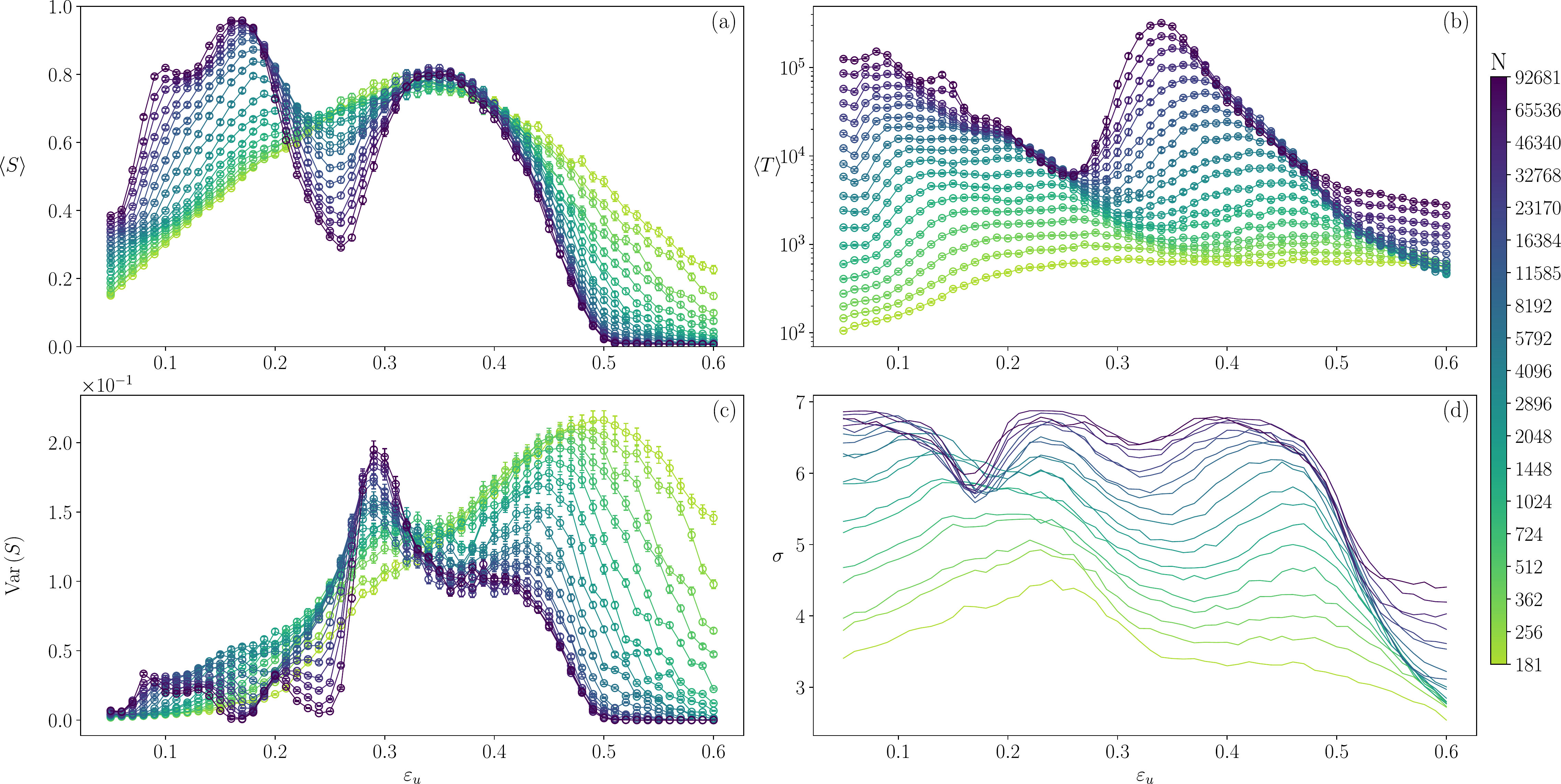}
	\caption{Metrics for the Barabási-Albert network of $\langle k \rangle=10$, $\varepsilon_l=0.05$. Average over 1000 realizations.
	(a) Average relative size of the largest cluster $\langle S \rangle$, (b) variance $\mathrm{Var}\left(S\right)$, (c) average convergence time $\langle T \rangle$, (d) largest cluster size entropy $\sigma$. }
	\label{metrics_BA}
\end{figure}

\section{Phase coexistence}

\subsection{Effect of topology}

This section presents the detail of size effects for the HK for the different topologies, for the largest system size available in each case.

\begin{figure}[h]
	\centering
	\includegraphics[width=\linewidth]{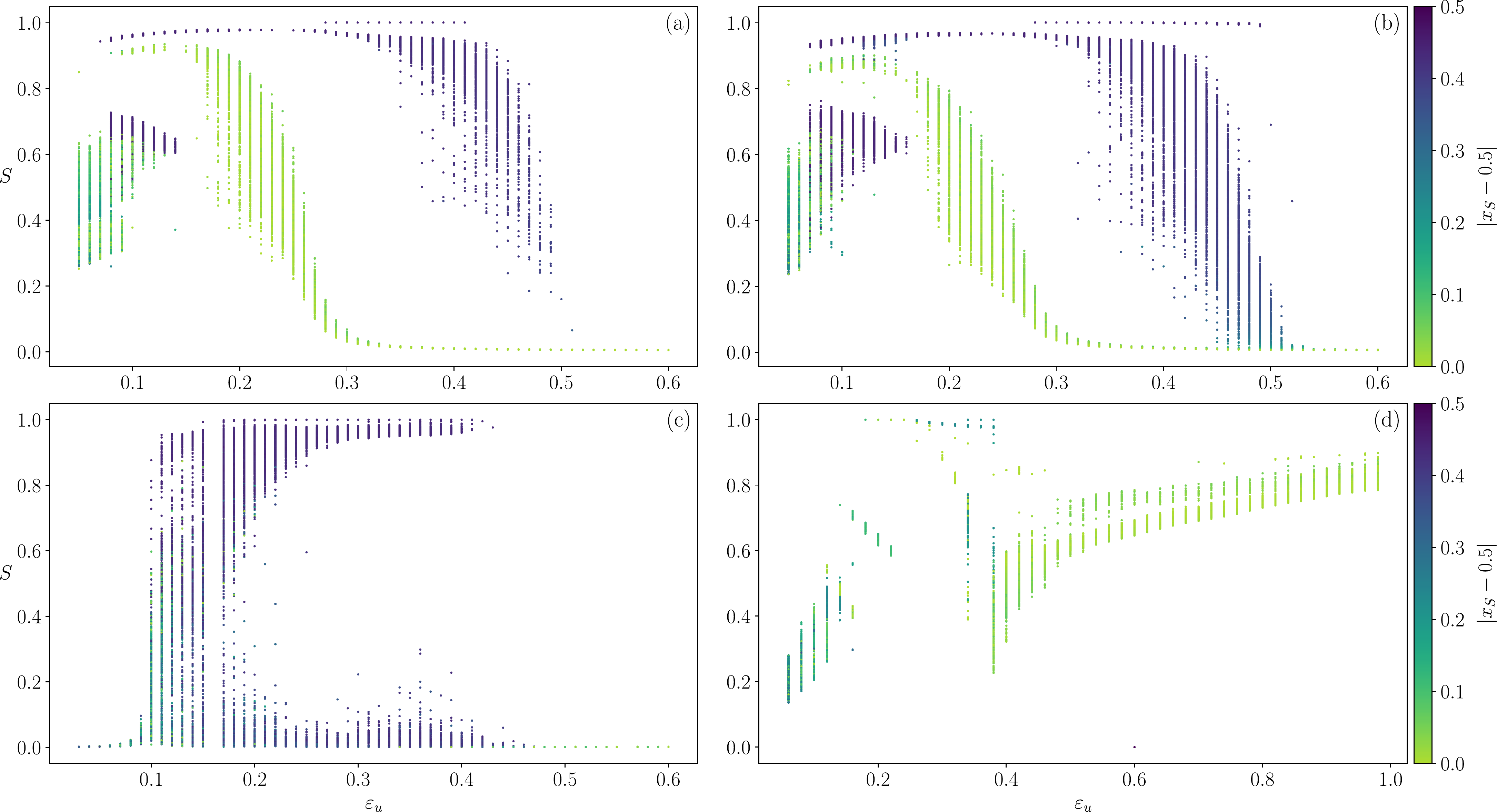}
	\caption{(a): Erdős–Rényi network of $\langle k \rangle=10$, $N=92681$, $\varepsilon_l=0.05$. (b): Barabási-Albert network of $\langle k \rangle=10$, $N=92681$, $\varepsilon_l=0.05$. (c): Square lattice including up to third neighbour interactions, $N = 16384$, $\varepsilon_l=0.03$. (d): Fully connected network (data taken with the permission of the authors from~\cite{schawe2020open}), $\varepsilon_l=0.05$.
	Note that due to the high occurrence of non-converging realizations at higher $N$, the scatter plot for the square lattice is not performed at the largest available size but at $N = 16384$.}
	\label{comparo_scatter_S}
\end{figure}

\subsection{Erd\H{o}s-R\'{e}nyi (ER)}

This section presents the videos of the full evolution of $x_i(t)$ as a function of $x_i(0)$ 
leading to the snapshots of Fig.7 of the main text.

\begin{itemize}
    \item  ER network with $\langle k \rangle$=10, $\varepsilon_l=0.05$,  $\varepsilon_u=0.35$, $N=92681$. Weak consensus, mild opinion (\textit{"Mild"}) phase.\\
    \href{https://zenodo.org/record/7455641/files/scatter_x0_xt_ER_k%3D10_N%3D92681_mild.mp4?download=1}{\texttt{scatter\_x0\_xt\_ER\_k=10\_N=92681\_mild.mp4}}\\
    
    \item ER network with $\langle k \rangle$=10, $\varepsilon_l=0.05$,  $\varepsilon_u=0.35$, $N=92681$. Consensus extreme opinion (\textit{"Skewed"}) phase.\\
    \href{https://zenodo.org/record/7455641/files/scatter_x0_xt_ER_k%3D10_N%3D92681_skewed.mp4?download=1}{\texttt{scatter\_x0\_xt\_ER\_k=10\_N=92681\_skewed.mp4}}\\
    
    \item ER network with $\langle k \rangle$=10, $\varepsilon_l=0.05$,  $\varepsilon_u=0.35$, $N=92681$. Unanimity extreme opinion (\textit{"U-turn"}) phase.\\
    \href{https://zenodo.org/record/7455641/files/scatter_x0_xt_ER_k%3D10_N%3D92681_uturn.mp4?download=1}{\texttt{scatter\_x0\_xt\_ER\_k=10\_N=92681\_uturn.mp4}}\\
\end{itemize}

\subsection{Barab\'{a}si-Albert (BA)}

This section presents the plots corresponding to Fig.5, Fig.7, and Fig.8 of the main text, for the Barabási-Albert network.

\begin{figure}[h]
	\centering
     \includegraphics[width=\linewidth]{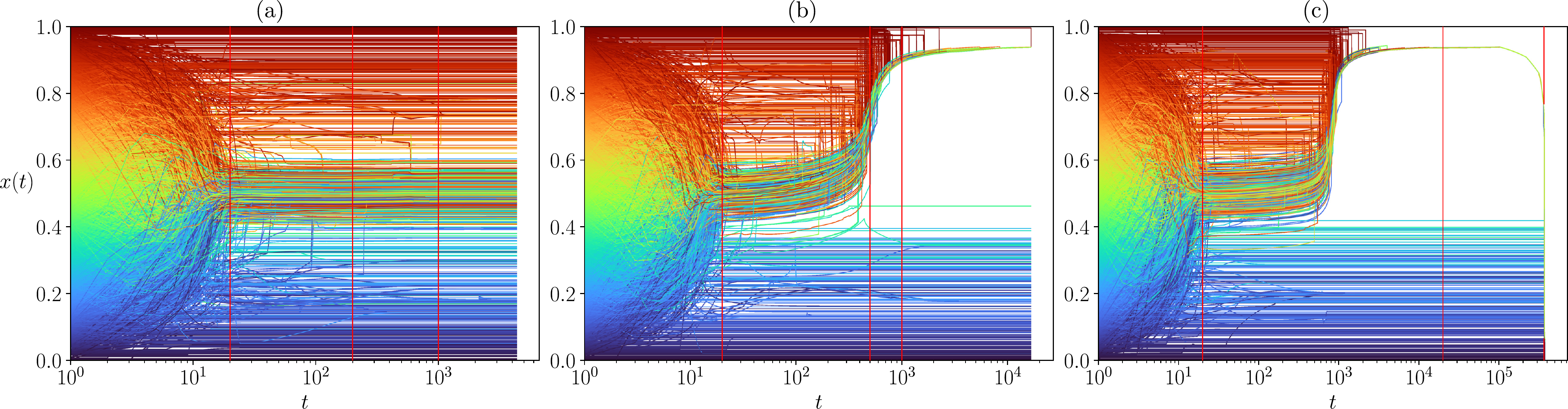}
	\caption{Example of the opinion evolution of the agents  for a society where the interactions are constrained by a Barabási-Albert network of $\langle k \rangle = 10$, and with confidence interval $\varepsilon_l = 0.05$, $\varepsilon_u = 0.35$, $N=92681$. The vertical lines indicate the time at which   snapshots shown in  Fig.~\ref{opinion_evolution_confidence} were taken. (a) Mild opinion phase: The agents located around the central opinion do not meet the criterion ($\delta x < 10^{-3}$) to constitute a single cluster. (b) Skewed phase. The final state is a dense strand containing the majority of the society while a few agents remain isolated. (c) U-turn phase: The opinion of the society is overturned from one extreme to the opposite one, the whole society is involved. Note that for readability and computational cost, only a set of 4000  agents taken at random is displayed.}
	\label{trajectories}
\end{figure}

\begin{figure}[h]
	\centering
    \includegraphics[width=\linewidth]{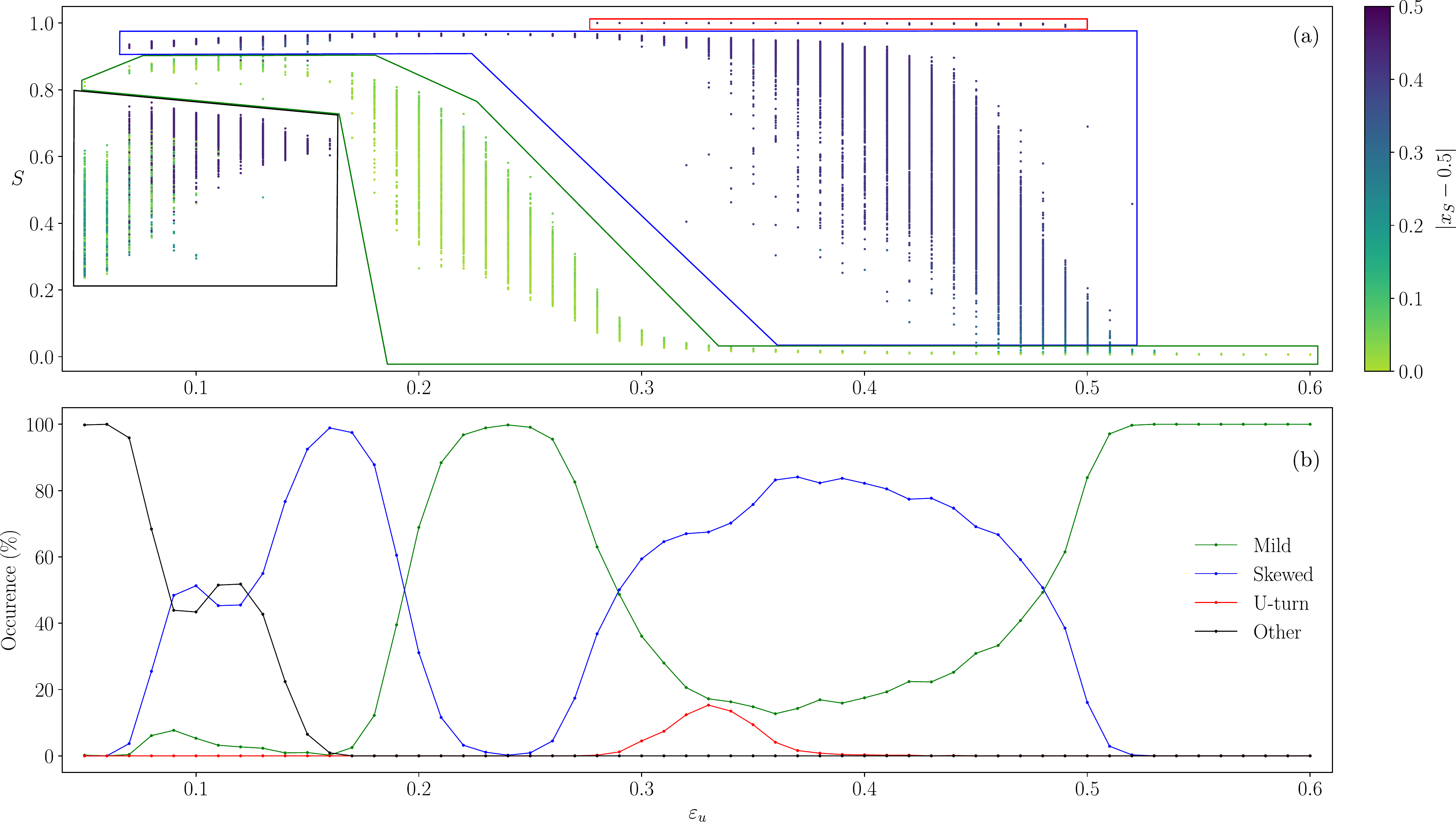}
	\caption{Probability of each type of phase for societies having  $\varepsilon_l = 0.05$, as a function of $\varepsilon_u$, for $N=92681$ and  interactions  constrained by a Barabási-Albert network of $\langle k \rangle = 10$. Averages calculated over 1000 samples. (a) The coloured boxes represent the samples used to compute the percentages shown in panel(b). (b) probability of occurrence of each of the three phases observed in the steady state.  }
	\label{occurence_rate}
\end{figure}

\begin{figure}[H]
	\centering
    \includegraphics[width=\linewidth]{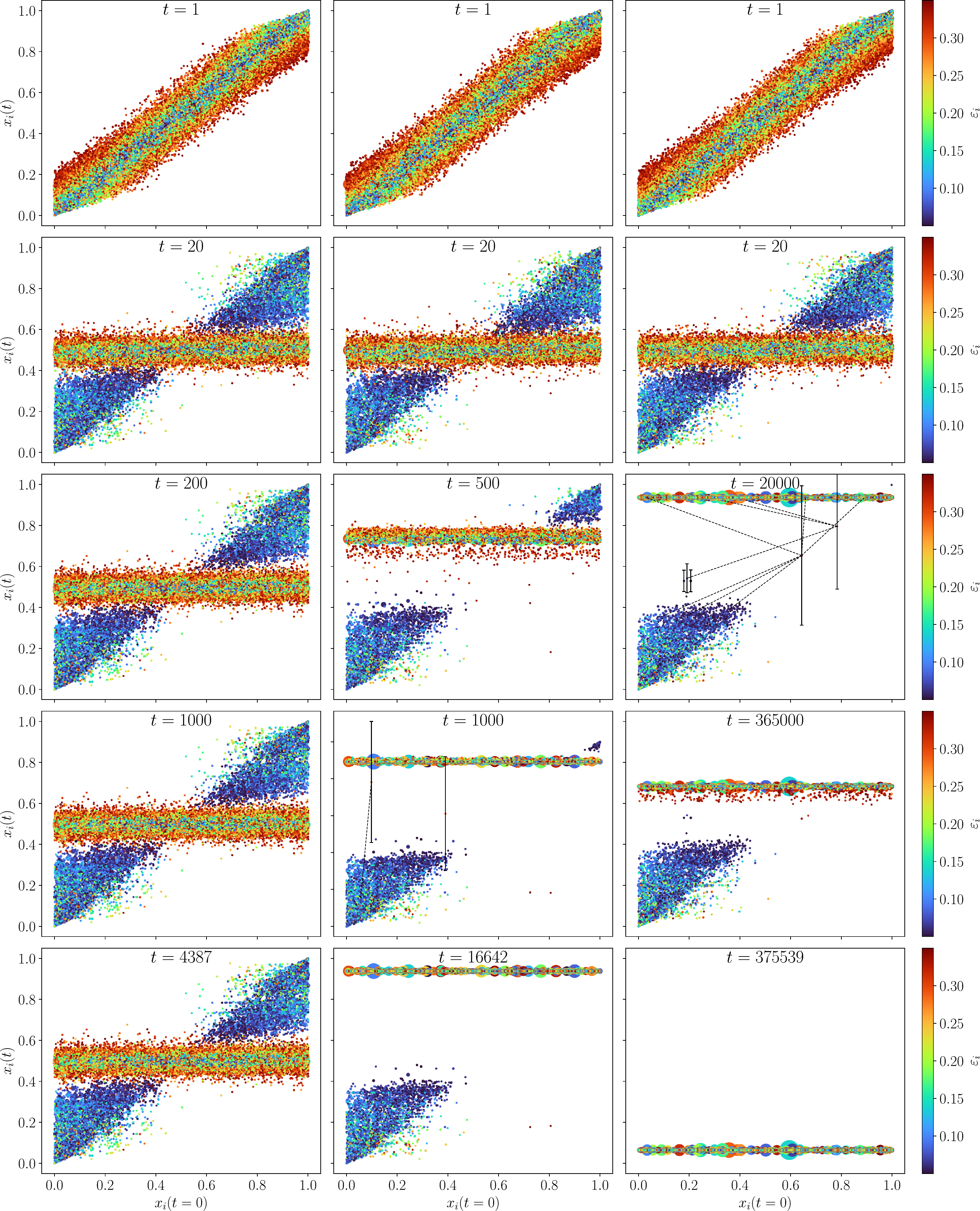}
	\caption{ Snapshots of the evolution of the agents opinion as a function of their initial opinion, with the size of the dots proportional to the degree, color-coded by their confidence for a society of $N=92681$ agents where the interactions are constrained by a Barabási-Albert of $\langle k \rangle = 10$, and with confidence interval $\varepsilon_l = 0.05$, $\varepsilon_u = 0.35$. Notice that the time to final convergence is increasing from left to right.  Left column:  Mmild opinion phase. Middle column: Skewed phase. Right column: U-turn phase . Bridges are highlighted in some frames: solid black vertical lines represent the confidence interval of the agents and dotted lines represent \textit{active links} (i.e. agents are topologically connected and within confidence reach).}
	\label{opinion_evolution_confidence}
\end{figure}

The videos of the full evolution of agents opinion as a function of their initial opinion, the snapshots of  Fig.~\ref{opinion_evolution_confidence} are taken from them:

\begin{itemize}
    \item  BA network with $\langle k \rangle$=10, $\varepsilon_l=0.05$,  $\varepsilon_u=0.35$, $N=92681$. Weak consensus, mild opinion (\textit{"Mild"}) phase.\\
    \href{https://zenodo.org/record/7455641/files/scatter_x0_xt_BA_k%3D10_N%3D92681_mild.mp4?download=1}{\texttt{scatter\_x0\_xt\_BA\_k=10\_N=92681\_mild.mp4}}\\
    
    \item BA network with $\langle k \rangle$=10, $\varepsilon_l=0.05$,  $\varepsilon_u=0.35$, $N=92681$. Consensus extreme opinion (\textit{"Skewed"}) phase.\\
    \href{https://zenodo.org/record/7455641/files/scatter_x0_xt_BA_k%3D10_N%3D92681_skewed.mp4?download=1}{\texttt{scatter\_x0\_xt\_BA\_k=10\_N=92681\_skewed.mp4}}\\
    
    \item BA network with $\langle k \rangle$=10, $\varepsilon_l=0.05$,  $\varepsilon_u=0.35$, $N=92681$. Unanimity extreme opinion (\textit{"U-turn"}) phase.\\
    \href{https://zenodo.org/record/7455641/files/scatter_x0_xt_BA_k%3D10_N%3D92681_uturn.mp4?download=1}{\texttt{scatter\_x0\_xt\_BA\_k=10\_N=92681\_uturn.mp4}}\\
\end{itemize}

\subsection{Square Lattice with third degree neighbours}

This section presents the videos for the trajectories shown in Fig.11 in the main text, for the Square Lattice with third degree neighbours. Because of the spatial embedding, these trajectories can be visualized in 2D.

\begin{itemize}
    \item  Square Lattice with $\langle k \rangle = k = 12$, $\varepsilon_l=0.03$,  $\varepsilon_u=0.35$, $N=163884$. Weak consensus, mild opinion (\textit{"Mild"}) phase.\\
    \href{https://zenodo.org/record/7455641/files/traj_2D_SL_k%3D12_N%3D16384_mild.mp4?download=1}{\texttt{traj\_2D\_SL\_k=12\_N=16384\_mild.mp4}}\\
    
    \item  Square Lattice with $\langle k \rangle = k = 12$, $\varepsilon_l=0.03$,  $\varepsilon_u=0.35$, $N=163884$.  Consensus extreme opinion (\textit{"Skewed"}) phase.\\
    \href{https://zenodo.org/record/7455641/files/traj_2D_SL_k%3D12_N%3D16384_skewed.mp4?download=1}{\texttt{traj\_2D\_SL\_k=12\_N=16384\_skewed.mp4}}\\
    
    \item  Square Lattice with $\langle k \rangle = k = 12$, $\varepsilon_l=0.03$,  $\varepsilon_u=0.38$, $N=163884$.  Unanimity extreme opinion (\textit{"U-turn"}) phase.\\
    \href{https://zenodo.org/record/7455641/files/traj_2D_SL_k%3D12_N%3D16384_uturn.mp4?download=1}{\texttt{traj\_2D\_SL\_k=12\_N=16384\_uturn.mp4}}\\
\end{itemize}

\bibliography{lit}